\documentclass[%
 reprint,
nofootinbib,
 amsmath,amssymb,
 aps,showkeys,
]{revtex4-1}

\usepackage{graphicx}% Include figure files
\usepackage{dcolumn}% Align table columns on decimal point
\usepackage{bm}% bold math
\usepackage{textpos}
\usepackage{hyperref}
\makeatletter
\let\@fnsymbol\@arabic
\makeatother

\begin{document}

% The following information is for internal review, please remove them for submission
\widetext
%\leftline{Submitted: 20 Dec 2015}
%\leftline{Accepted:}
%\leftline{Published:}

\preprint{APS/123-QED}

\title{Meteorites and the RNA World: A Thermodynamic Model of Nucleobase Synthesis within Planetesimals}% Force line breaks with \\
%\thanks{Submitted to Astrobiology}%

\author{Ben K. D. Pearce\textsuperscript{1,2,5}}
\author{Ralph E. Pudritz\textsuperscript{1,2,3,4,6}}%

\begin{textblock*}{100mm}(-0.35cm,-1.1cm)
Submitted: 20 Dec 2015
\end{textblock*}

\begin{textblock*}{100mm}(.37\textwidth,-1.1cm)
Accepted: 1 July 2016
\end{textblock*}

\begin{textblock*}{100mm}(.77\textwidth,-1.1cm)
Published: 21 Nov 2016
\end{textblock*}

\begin{textblock*}{100mm}(.26\textwidth,-0.7cm)
Astrobiology (2016) Vol. 16, Issue 11, 853--872
\end{textblock*}

\begin{textblock*}{100mm}(.75\textwidth,-0.7cm)
doi: \href{http://dx.doi.org/10.1089/ast.2015.1451}{10.1089/ast.2015.1451}
\end{textblock*}


\begin{abstract}
The possible meteorite parent body origin of Earth's pregenetic nucleobases is substantiated by the guanine (G), adenine (A) and uracil (U) measured in various meteorites. Cytosine (C) and thymine (T) however are absent in meteorites, making the emergence of a RNA and later RNA/DNA/protein world problematic. We investigate the meteorite parent body (planetesimal) origin of all nucleobases by computationally modeling 18 reactions that potentially contribute to nucleobase formation in such environments. Out of this list, we identify the two most important reactions for each nucleobase and find that these involve small molecules such as HCN, CO, NH$_3$, and water that ultimately arise from the protoplanetary disks in which planetesimals are built. The primary result of this study is that cytosine is unlikely to persist within meteorite parent bodies due to aqueous deamination. Thymine has a thermodynamically favourable reaction pathway from uracil, formaldehyde and formic acid, but likely did not persist within planetesimals containing H$_2$O$_2$ due to an oxidation reaction with this molecule. Finally, while FT synthesis is found to be the dominant source of nucleobases within our model planetesimal, NC synthesis may still be significant under certain chemical conditions (e.g. within CR2 parent bodies). We discuss several major consequences of our results for the origin of the RNA world.
\end{abstract}

%Use showkeys class option if keyword display desired              
\maketitle

\footnotetext[1]{Origins Institute, McMaster University, ABB 241, 1280 Main St, Hamilton, ON, L8S 4M1, Canada}
\footnotetext[2]{Department of Physics and Astronomy, McMaster University, ABB 241, 1280 Main St, Hamilton, ON, L8S 4M1, Canada}
\footnotetext[3]{Center for Astronomy, University of Heidelberg, Institute for Theoretical Astrophysics, Albert-Ueberle-Str. 2, 69120 Heidelberg, Germany}
\footnotetext[4]{Max Planck Institute for Astronomy, Königstuhl 17, 69117 Heidelberg, Germany}
\footnotetext[5]{E-mail: pearcbe@mcmaster.ca}
\footnotetext[6]{E-mail: pudritz@physics.mcmaster.ca}

\section{Introduction}

After the late heavy bombardment on the early Earth \citep{2005Natur.435..466G}, the rate of meteorite and comet impacts dropped low enough for the planet to cool and sustain liquid water. At that time, it is conceivable that meteorites, comets and interplanetary dust particles (IDPs) could have exogenously delivered a variety of important biomolecules---including nucleobases. The possibility that nucleobases could react, in early planetary environments, with sugars (ribose and deoxyribose) and phosphate to build the nucleotides that polymerized into RNA and DNA has driven a substantial body of biochemical research for decades \citep{1961Natur.191.1193O,1966Sci...154..784S,Reference10,1968GeCoA..32..175H,Reference53,Reference34,Reference20,2008OLEB...38..383L,2011OLEB...41..437S}.    
A data-storing molecule with an ability to make imperfect copies of itself, and the ability to form through purely natural means, is a necessary first requirement for life to emerge. On the prebiotic Earth, this molecule has long been thought to be RNA \citep{2013AsBio..13..391N}, due to its ability to both store and replicate data, and catalyze chemical reactions (such as catalyzing self-replication). It has been shown that RNA can be synthesized abiotically through the polymerization of ribonucleotides on specific clays \citep{2004AdSpR..33..100F}, in the presence of metal ion catalysts \citep{Reference58}, as well as in lipid bilayers \citep{Reference59}. 

The possibility of a meteorite parent body origin of the primeval Earth's nucleobases is substantiated by the G, A and U found in several carbonaceous chondrites on Earth \citep{2011PNAS..10813995C,2008E&PSL.270..130M,Reference37,1981GeCoA..45..563S,1979Natur.282..709S,1977GeCoA..41..961V,1975GeCoA..39..471H,1968GeCoA..32..175H,1964Sci...146.1291H}. 
(Carbonaceous chondrites are a meteorite type known for having high water and organic contents.) These nucleobases are thought to be extraterrestrial in origin, though it is not well understood which reactions are responsible for their synthesis within the parent bodies.

One of the major questions this hypothesis faces, is why hasn't C or T been measured in any carbonaceous chondrites? If the first self-replicating molecules on the early Earth were RNA molecules \citep{2013AsBio..13..391N}, then in order for the planetesimal and cometary origin hypothesis for the Earth's pregenetic nucleobases to be complete, then C---which base pairs with G in RNA---would have had to been available from meteorites. Furthermore, in order for the first (more stable) DNA molecules to form and potentially replace RNA as the main data-storing molecule, T---which base pairs with A in DNA---would have also had to been present. Several sources for  the origin of T other than meteorites might be possible, including interplanetary dust particles \citep{2014ApJ...793..125N}. We point out however, that the incorporation of T into DNA may have been a result of evolutionary tinkering, arising from the methylation of U by thymidylate synthase (an enzyme that converts U nucleotides into T nucleotides) \citep{Reference68}.

It also should be noted that although nucleobases were delivered to the primeval Earth, the reactions that produced the first nucleotides may not have necessarily used nucleobases as reactants. One of the challenges is that ribose is difficult to form \citep{Reference60,Reference61} and the addition of ribose to pyrimidines has proven elusive \citep{Reference58}, as has recently been emphasized by \citet{Reference56}.  \citet{Reference56} have demonstrated that the activated pyrimidine ribonucleotides (cytidine and uridine monophosphate) can be synthesized from reactions involving cyanamide, cyanoacetylene, glycolaldehyde, glyceraldehyde, inorganic phosphate and UV radiation. Interestingly, two of these reactants (glycolaldehyde and cyanoacetylene) have been recently detected in cometary materials \citep{2015SciA....115863B,2011ARA&A..49..471M}. The results from \citet{Reference56} could explain how the first chains of RNA possibly formed without C present in primordial meteorites. The activated purine nucleotides on the other hand still haven't been successfully synthesized in the lab, however scientists have recently gotten very close. The adenosine nucleoside has been synthesized by reacting its two constituents (adenine and ribose) \citep{1963Natur.199..222P,Reference57}, and the adenosine nucleoside has now been phosphorylated with the mineral schreibersite to form the unactivated (5') adenosine nucleotide \citep{Reference64}.

This paper investigates a protoplanetary disk origin for nucleobases, where adenine (A), guanine (G), cytosine (C), thymine (T) and uracil (U) were synthesized within planetesimals and comets, and delivered to the primordial Earth by the fragments of these bodies: meteorites, comets and interplanetary dust particles. (Planetesimals are 1--100 km-diameter rocky and/or icy bodies originating from the protoplanetary disk.)

As part of a long term project to understand the formation of biomolecules and their delivery to forming planets, \citet{2014ApJ...783..140C} first collated and analyzed abundances of amino acids in carbonaceous chondrites. Theoretical work on the origin of amino acids by means of aqueous Strecker reactions occurring within planetesimals was then carried out and compared with the meteoritic record \citep{Reference44}.

We extended this approach to nucleobases by first presenting the available data on nucleobase abundances within meteorites \citep{Reference46}. We then performed an extensive survey of the most frequently discussed chemical methods that have been employed or suggested as pathways for the abiotic formation of nucleobases. This survey was presented as a starting point in order to understand the reaction pathways that could occur within planetesimals. The comprehensive list was reduced by disregarding reactions that are unlikely to occur within planetesimals. A final list of 15 candidate nucleobase reaction pathways within planetesimals was then proposed \citep{Reference46} and we have since added three additional candidate reactions that were missed in the original survey (see Table~\ref{Candidates} for the chemical reaction equations).

In this paper, we computationally model the proposed 18 candidate nucleobase synthesis reactions within planetesimals using a chemical equilibrium software package called ChemApp. Our main goal is to give a theoretical explanation of the abundances and relative frequencies of nucleobases as observed in meteorites. Most importantly, we demonstrate why planetesimal conditions do not give rise to C and T and begin to discuss the alternatives.

First, in section~\ref{theorysec}, we outline the candidate reaction pathways for nucleobase synthesis within planetesimals, the theory and assumptions behind our thermodynamic model, and the varying stabilities of nucleobases in aqueous solution. Next, in section~\ref{methodssec}, we outline the computational methods and planetesimal environmental conditions. In section~\ref{resultssec} we present and analyze the results from our chemical equilibrium simulations. Then in section~\ref{discussionsec}, we discuss the implications of our simulations on the origin of the RNA world, investigate the main driver of nucleobase synthesis within planetesimals, explain discrepancies between our simulation abundances and the meteoritic record, and summarize the most important reactions for nucleobase synthesis within planetesimals. Finally, in section~\ref{conclusionssec}, we summarize the main results and conclusions of this work. In the appendices, we validate our substitute reactant for CA synthesis, and discuss the caveats in simulating competition between reactants with chemical equilibrium models.

%Nucleobase reaction simulations are run in groups, and individually, to understand how abundantly they can produce while competing and not competing with other reactions.

\section{Theory}\label{theorysec}

\subsection{Candidate Reaction Pathways}

Out of the most discussed abiotic nucleobase reaction mechanisms and pathways discussed in the literature, \citet{Reference46} argued that 15 reactions are potential contributers to the nucleobases synthesized within the parent bodies of meteorites. In this paper, we add three more reactions to this list for a total of 18 candidate reactions. These 18 candidate reactions are separated into three types: Fischer-Tropsch (FT), non-catalytic (NC) or catalytic (CA). The commonly discussed FT reactions \citep{1974OrLi....5...57A} involve gaseous ammonia, carbon monoxide and hydrogen in the presence of a catalyst such as alumina or silica. NC reactions are categorized based on their lack of a required catalyst, and CA reactions encapsulate the remaining, non-FT catalytic reactions. It was previously suggested that the FT reaction mechanism best supports the meteoritic record of nucleobases \citep{Reference46}.

The candidate reactions were selected based on their ability to react in the environmental conditions within a planetesimal, and the availability of their reactants within comets. Reactant availability in comets was chosen as a requirement because comets are the most unmodified bodies in the solar system \citep{2008tnoc.book..165R} and the molecules in comets could have also been available to planetesimals at the time of the latter's formation \citep{2004M&PS...39.1577S,2011IAUS..280..288A}. Reactant availability in meteorites on the other hand was not a requirement, as most carbonaceous chondrite matrices are not thought to be pristine, being depleted in volatiles to varying degrees \citep{2005PNAS..10213755B}. Carbonaceous chondrites are also from parent bodies that have undergone significant aqueous alteration \citep{2014ApJ...783..140C}, and are for a number of reasons more susceptible to weathering than other meteorite types \citep{2006mess.book..853B}. Therefore we assert that cometary concentrations may be more likely to represent the molecular concentrations in planetesimals during the formation of the solar system.

Of the considered catalytic reactions for nucleobase synthesis within planetesimals (i.e. FT and CA), only the reactions whose catalysts were present in meteorites made it through to the candidate list. Many other abiotic nucleobase synthesis mechanisms were disregarded due to external energies required for synthesis that are unlikely to be found in a planetesimal interior, e.g. Miller-Urey experiments requiring a high voltage electric discharge to synthesize G and A. 

The three reactions that have been added to the candidate list since our previous paper \citep{Reference46} are: uracil synthesis from neat formamide in the presence of Murchison meteorite powder or titanium dioxide \citep{2011OLEB...41..437S,Reference52} (reaction 61), thymine synthesis from the aqueous reaction of uracil, formaldehyde and formic acid \citep{Reference53} (reaction 62), and thymine synthesis from neat formamide in the presence of titanium dioxide \citep{Reference52} (reaction 63).

The 18 total candidate reactions are listed in Table~\ref{Candidates}. The chemical equations are either directly copied from the proposed reaction pathway listed in the original study, or are formulated based on the reactants used in the laboratory experiments and the nucleobase they produced. In the case of FT reactions, liquid water is added as a potential product due to the fact that it generally forms along with nucleobases in the laboratory experiments \citep{1981coge.conf....1A}. For the deamination of C (reaction 32), where C reacts with liquid water to form U, ammonia is also added as a potential product in order to perfectly balance the reaction. Finally, for the CA synthesis of A (reaction 24), where neat formamide reacts to form A, \citet{ANIE:ANIE201108907} and \citet{doi:10.1021/jp409296k} suggest formamide dehydration to be the first reaction step, therefore we include H$_2$O as an additional product for this reaction.

For the catalytic reactions, the chemical equation includes catalysts written above the reaction arrow. The catalysts used in these reactions are alumina (Al$_2$O$_3$), silica (SiO$_2$), nickel-iron alloy (NiFe), titanium dioxide (TiO$_2$), and Murchison meteorite powder. These reactions were performed in the laboratory with sometimes several of these catalysts used together or separately, therefore a `$\|$' is used to signify `or' and a `+$\|$' is used to signify `and/or.'

\begin{table*}[t]
\centering
\caption{Candidate reaction pathways for nucleobase synthesis within meteorite parent bodies from \citet{Reference46}, \citet{2011OLEB...41..437S,Reference52} and \citet{Reference53}.\label{Candidates}}

\begin{tabular}{cccc}
\\
\multicolumn{1}{c}{No.} &
\multicolumn{1}{c}{Type} &
\multicolumn{1}{c}{Reaction} &
\multicolumn{1}{c}{Source(s)}\\ \hline
$\underline{Adenine}$ & \\
1 & FT & CO + H$_{2}$ + NH$_{3}$ $\xrightarrow{NiFe+||Al_{2}O_{3}+||SiO_{2}}$ A + H$_{2}$O & \citet{Reference42};\\
  & & & \citet{1968GeCoA..32..175H}\\
3 & NC & 5HCN$_{(aq)}$ $\rightarrow$ A$_{(aq)}$ & \citet{2008OLEB...38..383L}\\
4 & NC & HCN + NH$_{3}$ $\rightarrow$ A & \citet{Reference41};\\
 & & & \citet{ReferenceWaka}\\
6 & NC & 5CO + 5NH$_{3}$ $\rightarrow$ A + 5H$_{2}$0 & \citet{1968GeCoA..32..175H}\\
7 & NC & HCN + H$_{2}$0 $\rightarrow$ A & \citet{Reference12}\\
8 & NC & HCN + NH$_{3}$ + H$_{2}$0 $\rightarrow$ A & \citet{Reference27}\\
24 & CA & Formamide $\xrightarrow{Al_{2}O_{3}||SiO_{2}}$ A + H$_2$O & \citet{Reference31}\\
$\underline{Uracil}$ & \\[+1mm]
29 & NC & 2HCN$_{(aq)}$ + 2CH$_{2}$O$_{(aq)}$ $\rightarrow$ U$_{(aq)}$ + H$_{2(aq)}$ &  \citet{2008OLEB...38..383L}\\
32 & NC & C + H$_{2}$O $\rightarrow$ U + NH$_{3}$ & \citet{1995Natur.375..772R};\\
 & & & \citet{Reference13};\\
 & & & \citet{Reference10}\\
61 & CA & Formamide $\xrightarrow{Murchison||TiO_{2}}$ U & \citet{2011OLEB...41..437S};\\
 & & & \citet{Reference52}\\
$\underline{Cytosine}$ & \\
43 & FT & CO + H$_{2}$ + NH$_{3}$ $\xrightarrow{NiFe+||Al_{2}O_{3}+||SiO_{2}}$ C + H$_{2}$O & \citet{Reference42};\\
 & & & \citet{1968GeCoA..32..175H}\\
44 & NC & 3HCN$_{(aq)}$ + CH$_{2}$O$_{(aq)}$ $\rightarrow$ C$_{(aq)}$ & \citet{2008OLEB...38..383L}\\
49 & CA & Formamide $\xrightarrow{Al_{2}O_{3}||SiO_{2}}$ C & \citet{Reference31}\\
$\underline{Guanine}$ & \\
51 & FT & CO + H$_{2}$ + NH$_{3}$ $\xrightarrow{NiFe+||Al_{2}O_{3}+||SiO_{2}}$ G + H$_{2}$O & \citet{Reference42};\\
 & & & \citet{1968GeCoA..32..175H}\\
54 & NC & 5HCN$_{(aq)}$ + H$_{2}$O $\rightarrow$ G$_{(aq)}$ + H$_{2(aq)}$ & \citet{2008OLEB...38..383L}\\
$\underline{Thymine}$ & \\[+1mm]
58 & NC & 2HCN$_{(aq)}$ + 3CH$_{2}$O$_{(aq)}$ $\rightarrow$ T$_{(aq)}$ + H$_{2}$O & \citet{2008OLEB...38..383L}\\
62 & NC & U + CH$_{2}$O + Formic Acid + H$_{2}$O $\rightarrow$ T & \citet{Reference53}\\
63 & CA & Formamide $\xrightarrow{TiO_{2}}$ T & \citet{Reference52}\\
\hline
\multicolumn{4}{l}{\footnotesize NC: Non-catalytic, CA: Catalytic, FT: Fischer-Tropsch.}
\end{tabular}
\end{table*}

\subsection{Gibbs Free Energy of Formation}

There are three important Gibbs free energies to pay attention to when doing thermochemical calculations. The Gibbs free energy of formation, $\Delta G_f$, The Gibbs free energy of reaction, $\Delta G_r$, and the total Gibbs free energy of the system, $\Delta G$. Every molecule has a Gibbs free energy of formation which varies with temperature and pressure. It is an extensive quantity that essentially represents each molecule's formation favourability. The lower the value of $\Delta G_f$, the more easily the molecule will form, as it requires less free energy input. If $\Delta G_f$ is negative, the molecule should form spontaneously (given the necessary reactants are available). 

$\Delta G_f$ functions for all chemical species can be calculated by fitting their $\Delta G_f$ data to the function

\begin{equation} \label{coeff}
  G_f(T,P) = a + bT + cTln(T) + dT^2 + eT^3 + f/T + gP.
 \end{equation}
 
The Gibbs coefficients a--g are the requisite input for the equilibrium chemistry software used for our chemical reaction simulations.

For a chemical reaction to be thermodynamically favourable, it must have a negative $\Delta G_r$. $\Delta G_r$ is the Gibbs free energy of reaction and is calculated with the equation

\begin{equation}\label{freereaction}
\Delta G_r = \Sigma G^{products}_f - \Sigma G^{reactants}_f.
\end{equation}

$\Delta G_r$ must be negative to be favourable, because a reaction with a positive $\Delta G_r$ requires input energy, and will increase the total Gibbs free energy of the system.

When a system has reached equilibrium, the chemical reactions will have essentially ceased, as  there is no longer a series of reactions that can occur given the present concentrations that will result in a negative $\Delta G_r$. This underlines the complete concept behind thermodynamic chemical reaction simulations, which is to set the initial concentrations of reactant molecules in the system, and then calculate the resultant reactant and product concentrations which minimize $\Delta G$. The total Gibbs free energy of the system can be calculated by summing every molecule in the system's Gibbs free energy of formation,

\begin{equation}\label{totalGibbs}
\Delta G = \Sigma G^{all}_f.
\end{equation}

Catalysts do not play a role in the minimization of Gibbs free energy calculations, as catalysts do not contribute molecules to reactions. Catalysts only speed up the reaction time by lowering the activation energy---a variable that is not used in equilibrium calculations.  

\subsection{Model Assumptions}

In order for a planetesimal to reach chemical equilibrium---a primary assumption in using thermodynamic models---the planetesimal must offer a stable environment for the duration that the reactions can occur. This environmental stability is defined as the ability of reactants to remain in the phase in which they react (for the chemical reactions considered). Because meteorite parent bodies are thought to have temperatures providing aqueous interiors for timescales longer than 1 Myr \citep{2005E&PSL.240..234T}, all the reactants in Table~\ref{Candidates} should remain in the phase in which they react for at least this long.

HCN has a half-life in an aqueous solution of no longer than ten thousand years \citep{1984AdSpR...4...69P}. This means that all of the NC reactions requiring HCN (nos. 3, 4, 7, 8, 29, 44, 54 and 58) will finish occurring long before the planetesimal ceases to be aqueous. Similarly, the timescale for the deamination of cytosine into uracil (no. 32) is $\sim$ 17,000 years at 0$^{\circ}$C, reacting even faster with increasing temperatures. Thus reaction 32 will also finish occurring long before the planetesimal ends its aqueous lifespan. It must be noted that HCN has been measured in the Murchison meteorite \citep{2012ApJ...754L..27P}, which could suggest that the HCN-based reactions perhaps did not reach completion within the Murchison parent body. However, due to the release of HCN upon acidification of meteorite extracts, it has been suggested that the measured HCN was not a free reactant, but rather was tied up in --CN salts that formed from reactions with Fe$^{2+}$, Mg$^{2+}$, and Ca$^{2+}$ cations during the planetesimal's aqueous phase \citep{2012ApJ...754L..27P}.

Some nucleobases were produced in 2--288 hours from gaseous reactants (nos. 1, 43, 51, and 6) \citep{1968GeCoA..32..175H}. This is an extremely short timescale compared to the timescale a planetesimal will remain aqueous, therefore these reactions should finish long before the planetesimal interior cools and potentially traps the remaining reactants in ice. Although the limiting reagent for these four reactions, NH$_3$, is found within carbonaceous chondrites \citep{2011GeCoA..75.7585M,2009GeCoA..73.2150P}, it is suggested that these reactions still reached completion within their parent bodies by being completely depleted of their only carbon source (CO). This is discussed in greater detail in Section~\ref{ammonia}. Finally, the four CA reactions (nos. 24, 49, 61 and 63) were synthesized from a formamide solution in 48 hours \citep{2011OLEB...41..437S,Reference52,Reference31} and the NC thymine reaction (no. 62) was synthesized in  2--28 days \citep{Reference53}. These are also short reaction times in comparison to the planetesimal's aqueous lifetime, thus we conclude that all the reactions in Table~\ref{Candidates} can sufficiently be compared under the assumption of chemical equilibrium. 

It should be noted that formaldehyde, the limiting reagent of reaction 62 also has measured abundances in carbonaceous chondrites \citep{2011GeCoA..75.7585M,2009GeCoA..73.2150P}. However, it is thought that this measured formaldehyde is not free formaldehyde which would have been available for reaction. Instead, \citet{2009GeCoA..73.2150P} note that due to poor aldehyde/ketone extraction at high temperatures (80--100$^{\circ}$C), and poor high-temperature extraction of other, more soluble carbonyls, that the formaldehyde measured in meteorites is likely tied up in reversible bonds with other organic compounds, or chemically adsorbed onto clays.

The FT and CA reactions require a catalyst in order to synthesize nucleobases, thus the required catalysts (Al$_{2}$O$_{3}$, SiO$_{2}$, NiFe, TiO$_{2}$ and/or Murchison minerals) are assumed to be present in the simulated carbonaceous chondrite parent body environment. This assumption is validated by the presence of these minerals within carbonaceous chondrites \citep{1990Metic..25..323J,1956GeCoA...9..279W}.

We assume weak coupling for our chemical reaction simulations by only including the known reactants and the products of interest (usually an individual nucleobase) in each reaction simulation. This is a safe assumption when reactants are in much greater concentrations than the products in the simulated reaction environment \citep{Reference44}. Weak coupling simulations for amino acid synthesis within meteorite parent bodies have been demonstrated to produce relative amino acid abundances that well represent the relative abundances of amino acids in meteorites \citep{Reference44}. Conversely, attempts to simulate amino acid synthesis within meteorite parent bodies by including all potential amino acids as potential products in a single simulation results in several expected amino acids being completely unproductive. (Since there are a myriad of possible products that could be produced from a given solution, it is more accurate to simplify and simulate a single reaction than to increase the complexity and attempt to simulate all possible reactions at once.)

The final assumption of our thermodynamic model is that the simulation environment (the planetesimal) is an isolated thermodynamic system. This assumption is required because the simulation software assumes no exchange of particles or heat with the reservoir (in this case, the vacuum of space). Since the interiors of planetesimals are thought to generate heat from $^{26}$Al decay for a few million years \citep[and references therein]{2002aste.conf..559M}, planetesimals will maintain a pseudo-equilibrium between the heat generated from radionuclide decay and the heat lost from thermal emission during this period. This simplified model of a planetesimal is sufficient to obtain useful comparisons between reactions and the meteoritic data.

\subsection{Equilibrium Chemistry Software}\label{chemapp}

The computational nucleobase synthesis simulations are performed using a thermochemistry software library called ChemApp (distributed by GTT Technologies, http://gtt.mch.rwth-aachen.de/gtt-web/). The ChemApp subroutines are called in a program written by the authors using the FORTRAN language. This program requires the Gibbs coefficients from Equation~\ref{coeff} for each reactant and product, as well as their initial abundances, and the temperature and pressure of the system. The ChemApp library has also been used by \citet{Reference44} to run chemical equilibrium calculations in the simulation of amino acid synthesis.

To compute the reactant and product abundances for each reaction at equilibrium, ChemApp first breaks down the initial molecular abundances of the reactants into their elemental abundances (carbon, hydrogen, oxygen and nitrogen). ChemApp then builds the system back up into the combination of reactant and product abundances that provides the minimum value of $\Delta$G (Equation~\ref{totalGibbs}).

The Gibbs data used by the ChemApp subroutines is obtained from the CHNOSZ thermodynamic database (version 1.0.3 (2014-01-12), authored by Jeffrey M. Dick, http://www.chnosz.net/).

\subsection{Molecular Stability}\label{stability}

By comparing the aqueous nucleobase decomposition rates \citep{Reference45} with the experimental nucleobase reaction rates across various temperatures, we can set the effective temperature boundaries for nucleobase formation within planetesimals.

If the decomposition rate of a nucleobase exceeds its reaction rate, the nucleobase will not have measurable yields at equilibrium. Table~\ref{ReactionRates} lists the approximate experimental reaction and decomposition rates for the non-theoretical reactions in Table~\ref{Candidates}.

\begin{table}[!ht]
\centering
\caption{Reaction and decomposition rates at various temperatures for the known reactions in Table~\ref{Candidates}. Reaction rates are taken as the experiment durations from the corresponding experiments, with the exception of reaction 32 which is taken from a nucleobase decomposition experiment \citep{Reference45}. Hydrolysis half-lives are retrieved from decomposition experiments \citep{Reference45}. The effects of pressure on nucleobase decomposition are small\label{ReactionRates}.}

\begin{tabular}{ccccc}
\\
\multicolumn{1}{c}{No.} &
\multicolumn{1}{c}{Type} &
\multicolumn{1}{c}{Temp. ($^{\circ}$C)} &
\multicolumn{1}{c}{Reaction Time} &
\multicolumn{1}{c}{Half-life}\\ \hline
$\underline{Adenine}$ & \\
1 & FT & 50 & 52 hours & 286 years\\
1 & FT & 60 & 68 hours & 81 years\\
1 & FT & 75 & 19 hours & 14 years\\
1 & FT & 165 & 16 hours & 40 hours\\
1 & FT & 350 & 0.5 hours & 14.5 seconds\\
4 & NC & 120 & 20 hours & 58 days\\
6 & NC & 500 & 16 hours & $<$ 1 second \\
7 & NC & 110 & 24 hours & 144 days\\
8 & NC & 23 & 19 days & 13,000 years\\
8 & NC & 70 & 120 hours & 25 years\\
8 & NC & 90 & 24 hours & 3 years\\
24 & CA & 160 & 48 hours & 58 hours\\
$\underline{Uracil}$ & \\[+1mm]
32 & NC & 0 & 17,000 years & $>$10$^6$ years\\
32 & NC & 50 & 15 years & 18,000 years\\
32 & NC & 100 & 19 days & 12 years\\
32 & NC & 165 & 3.5 hours & 4 days \\
61 & CA & 160 & 48 hours & 6.5 days \\
$\underline{Cytosine}$ & \\
43 & FT & 0 & - & 17,000 years\\
43 & FT & 50 & 52 hours & 11 years\\
43 & FT & 60 & 68 hours & 3 years\\
43 & FT & 75 & 19 hours & 227 days\\
43 & FT & 142 & 8.5 hours & 13.5 hours\\
43 & FT & 165 & 16 hours & 2.5 hours\\
43 & FT & 350 & 6 hours & 1.5 seconds\\
49 & CA & 160 & 48 hours & 3.5 hours\\
$\underline{Guanine}$ & \\
51 & FT & 50 & 52 hours & 339 years\\
51 & FT & 60 & 68 hours & 88 years\\
51 & FT & 75 & 19 hours & 13 years\\
51 & FT & 165 & 16 hours & 21 hours\\
51 & FT & 350 & 6 hours & 4 seconds\\
$\underline{Thymine}$ & \\[+1mm]
62 & NC & 50 & - & 90,000 years\\
62 & NC & 100 & 28 days & 58 years\\
62 & NC & 120 & 28 days & 5 years\\
62 & NC & 140 & 50 hours & 211 days\\
63 & CA & 160 & 48 hours & 29 days\\
63 & CA & 205 & - & 15 hours\\
\hline
\multicolumn{5}{l}{\footnotesize NC: Non-catalytic, CA: Catalytic, FT: Fischer-Tropsch.} \\
\multicolumn{5}{l}{\footnotesize Hydrolysis half-lives for each temperature are calculated} \\ \multicolumn{5}{l}{\footnotesize using the Arrhenius equations derived from experiment \citep{Reference45}.}
\end{tabular}
\end{table}

Cytosine (C) is the least stable nucleobase with a half-life due to hydrolysis of approximately 3.5 hours at 165$^{\circ}$C, 15 years at 50$^{\circ}$C and 17,000 years at 0$^{\circ}$C. In contrast, thymine (T) is the most stable with a half-life of 18 days at 165$^{\circ}$C, 90,000 years at 50$^{\circ}$C and $>$10$^6$ years at 0$^{\circ}$C. At temperatures less than 142$^{\circ}$C, all experimental nucleobase reaction rates are faster than their corresponding aqueous solution decomposition rates. At 165$^{\circ}$C, only C decomposes quicker than it reacts, but both adenine (A) and guanine's (G) reaction rates are nearing their decomposition rates. This puts an approximate upper boundary of nucleobase synthesis within planetesimals at 165$^{\circ}$C. This coincides very nicely with the temperatures of planetesimal interiors from 3D thermal evolution simulations which range from 0--180$^{\circ}$C (during the aqueous phase of a 100km body) \citep{2005E&PSL.240..234T}.

Since reaction 6 of A was only synthesized at a very high temperature (500$^{\circ}$C) in the laboratory \citep{1968GeCoA..32..175H}, it is probably unlikely that this reaction is occurring within planetesimals. Within a planetesimal of this temperature, A would decompose in under a second after coming in contact with water.

It should be noted that some additional stability may have been afforded by adenine, guanine, and uracil within planetesimals, as they can be incorporated into HCN polymers in aqueous solution \citep{Reference62}. Because HCN polymers require acid hydrolysis (e.g. with HCl) to release the bonds between linked compounds \citep{Reference69,Reference27} it is conceivable that the HCN polymer offers additional support against the degradation of its incorporated nucleobases.

\section{Computational Methods}\label{methodssec}

\subsection{Calculating Gibbs Free Energy Coefficients}

The Gibbs coefficients (Equation~\ref{coeff}) for each molecule are obtained by performing a least squares fit to the corresponding $\Delta G_f$ data obtained from CHNOSZ.

\begin{figure}[ht!]
\centering
\includegraphics[width=80mm]{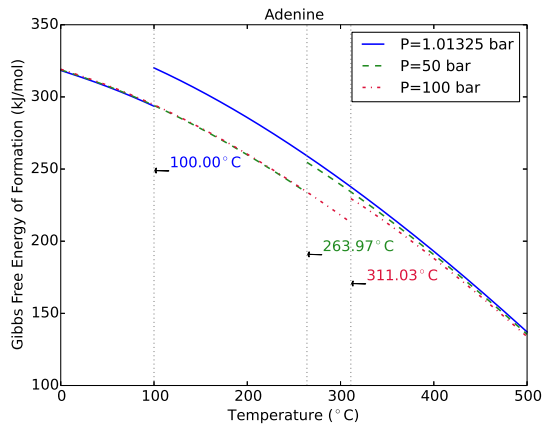}
\caption{The Gibbs free energy dependence on temperature and pressure for aqueous adenine. Temperature varies from 0$^{\circ}$C to 500$^{\circ}$C. The blue curve represents a pressure of 1.01325 bar, the green curve represents a pressure of 50 bar and the red curve represents a pressure of 100 bar.}
\label{GibbsAda}
\end{figure}

In Figure~\ref{GibbsAda} we illustrate the Gibbs data from CHNOSZ for aqueous A over three pressures and a range of temperatures. Notice how the Gibbs free energy curves are discontinuous at the boiling point of water for each pressure (1.01325 bar: 100$^{\circ}$C; 50 bar: 263.97$^{\circ}$C; 100 bar: 311.03$^{\circ}$C). The discontinuous increase in $\Delta$G$_f$ at the liquid-to-gas phase transition represents a decrease in thermodynamic favourability for the aqueous formation of A---as an increase in the $\Delta$G$_f$ of a product leads to a higher $\Delta$G$_r$ for its reaction (Equation~\ref{freereaction}).

It is important to note that the Gibbs free energies in Figure~\ref{GibbsAda} are practically independent of pressure for the temperatures below the liquid-to-gas phase transition. These three pressure curves differ by $<$ 1kJ from 0--100$^{\circ}$C. This lack of pressure dependence allows us to set the pressure of our thermodynamic system to a static 100 bar, making temperature and initial reactant concentrations the only dynamic simulation variables.

\begin{figure}[ht!]
\centering
\includegraphics[width=80mm]{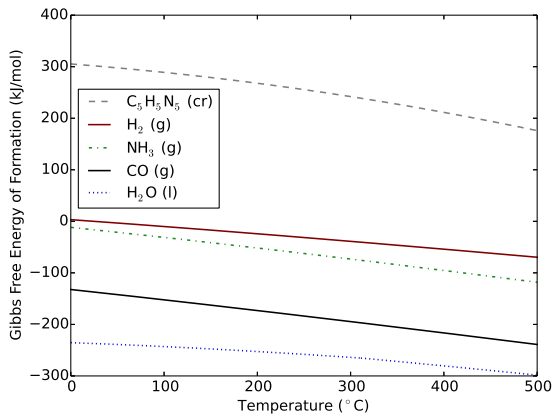}
\caption{The Gibbs free energies of the reactants and products of reaction 1 (the FT synthesis of adenine) at a pressure of 100 bar. Temperature varies from 0$^{\circ}$C to 500$^{\circ}$C. From top to bottom the curves represent adenine, H$_2$, NH$_3$, CO, and water.}
\label{GibbsR1}
\end{figure}

In Figure~\ref{GibbsR1} we illustrate the Gibbs free energies of formation for reaction 1: the FT synthesis of A (see Table~\ref{Candidates} for more detail). Notice how CO, NH$_3$ and H$_2$ are much more thermodynamically favourable than A, with practically all of their $\Delta$G$_f$ values being negative for the 0--500$^{\circ}$C range. It is obvious that A could not be synthesized from these three reactants---since they have lower, more favourable $\Delta$G$_f$ values---without also producing H$_2$O in the process. Water has the most thermodynamically favourable $\Delta$G$_f$ of all five molecules, making it slightly more favourable for the reactants to produce water and A than remain themselves. This underlines the importance of water being produced in FT synthesis, as the $\Delta G_r$ for reaction 1 without water would never be negative.

\begin{figure}[ht!]
\centering
\includegraphics[width=80mm]{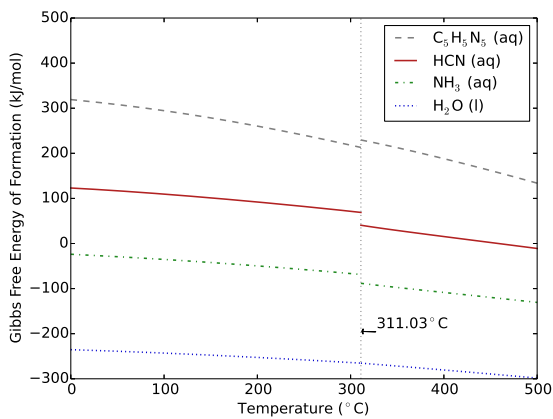}
\caption{The Gibbs free energies of the reactants and product of reaction 8 (the NC synthesis of adenine) at a pressure of 100 bar. Temperature varies from 0$^{\circ}$C to 500$^{\circ}$C. From top to bottom the curves represent adenine, HCN, NH$_3$, and water.}
\label{GibbsR8}
\end{figure}

Figure~\ref{GibbsR8} shows the $\Delta G_f$ values for reaction 8: the NC synthesis of A. The Gibbs free energy of A (C$_5$H$_5$N$_5$) is lower than five times the Gibbs free energy of HCN at every temperature. This makes it clear that this reaction should produce from a thermochemical standpoint, as it is more favourable to form A from five HCN molecules than it is for five HCN molecules to remain themselves. Although some authors suggest that the reaction pathway from HCN to A (C$_5$H$_5$N$_5$) may be more complicated than combining five HCN molecules (and could require NH$_3$ as an intermediate reactant and product) \citep{1961Natur.191.1193O}, since intermediate reactions don't effect the results at equilibrium, the NH$_3$ and H$_2$O abundances in the simulation of reaction 8 are not likely to change from their initial concentrations.

Unfortunately a limitation arises in the simulation of the CA reactions from Table~\ref{Candidates}, as the CHNOSZ database does not have any Gibbs free energy data for the formamide molecule. In order to study this reaction, we instead employ the closest molecule for which CHNOSZ has Gibbs data. The substitute molecule chosen is the carbamoyl functional group (-CONH$_2$), which is the side chain of the amino acid glutamine. Identically to formamide, the carbamoyl functional group has a C-NH$_2$ bond and a C=O double bond. Formamide (CH$_3$NO) just differs from the carbamoyl functional group by also having a single hydrogen atom bonded to the carbon atom. There is one other discrepency between formamide and the substitute molecule: CHNOSZ only has data for the carbamoyl functional group in an aqueous solution, yet the CA reactions are performed experimentally in a neat formamide solution (dissolved in itself). An estimate on the difference in Gibbs free energies of formation between liquid formamide and the aqueous carbamoyl functional group are detailed in Appendix A.

In Figure~\ref{GibbsR23} we illustrate the Gibbs free energies of formation for reaction 24 (the CA synthesis of A), with the formamide substitute molecule. Because the formamide substitute has a much lower Gibbs free energy of formation than the sum of the products, it is likely more favourable for the formamide substitute to remain itself than to form A and H$_2$O.

\begin{figure}[ht!]
\centering
\includegraphics[width=80mm]{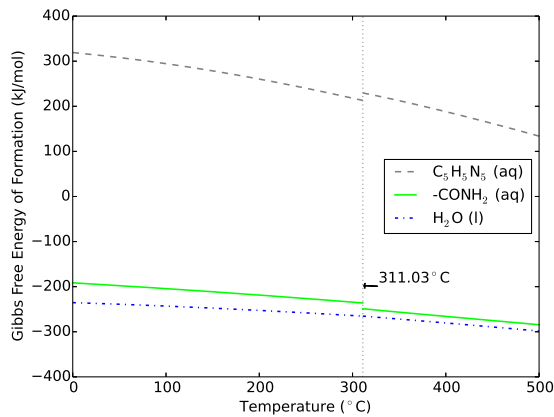}
\caption{The Gibbs free energies of the reactants and product of reaction 24 (the CA synthesis of adenine) at a pressure of 100 bar. Temperature varies from 0$^{\circ}$C to 500$^{\circ}$C. The gray curve (top) is adenine, the green curve (middle) represents the carbamoyl functional group (formamide Gibbs data unavailable) and the blue curve (bottom) represents water.}
\label{GibbsR23}
\end{figure}

As a final note, due to the lack of non-aqueous Gibbs free energy data for HCN in CHNOSZ, reaction 4 of A is simulated as an aqueous reaction even though \cite{Reference41} and \cite{ReferenceWaka} performed this reaction in the laboratory without water.

\subsection{Planetesimal Interiors}\label{interiors}

The temperature boundaries for the interior of a model carbonaceous chondrite parent body in a previous biomolecule simulation study were selected as 0--500$^{\circ}$C \citep{Reference44}. These values were based on models of the thermal evolution of planetesimals due to $^{26}$Al decay \citep{2005E&PSL.240..234T,2002aste.conf..559M}. In simulations by \citet{2005E&PSL.240..234T}, 100 km-diameter planetesimal interiors reached a maximum of 180$^{\circ}$C. Simulations by \citet{2002aste.conf..559M} produced similar results, with their smallest-radius planetesimal simulations reaching a maximum interior temperature of 227$^{\circ}$C. The temperatures upwards of 227$^{\circ}$C were selected by \citet{Reference44} to conform to various studies that classified temperature ranges within the parent bodies of the various carbonaceous chondrite subclasses and petrologic types \citep{Reference35,2013AREPS..41..529W}.

Our simulation temperatures conform with those from \citet{Reference44}, beginning at 0$^{\circ}$C (as none of our chemical reactions are solid state) and run to a maximum of 500$^{\circ}$C.

Our chosen static pressure of 100 bar should be within a few factors of the theoretical maximum for the interior of meteorite parent bodies. Using a central pressure $p \sim \frac{2}{3} \pi G \rho^2 R^2$, 100 bar would be slightly more than the maximum pressure within 19 Fortuna: the asteroid postulated to be the parent body source of CM meteorites \citep{2002aste.conf..653B}. Though, as previously shown in Figure~\ref{GibbsAda}, pressure will not play a large role in nucleobase synthesis while the planetesimal remains in the aqueous phase.

Our fiducial planetesimal model conforms to the fiducial model in a previous biomolecule simulation study \citep{Reference44}, which was chosen to match cases 1--3 of the modelled carbonaceous chondrite parent bodies proposed in a numerical thermal evolution study by \citet{2005E&PSL.240..234T}. This model is a spherical rock with a porosity of 20$\%$, a radius of 50 km, a rock density of 3000 kg m$^{-3}$ and an ice water of density 917 kg m$^{-3}$ completely filling the pores of the body. Our fiducial model is also consistent with recent work by \citet{2016arXiv160305979L}, who performed an extensive suite of 2D and 3D numerical thermo-mechanical evolution simulations covering various planetesimal radii, formation times, and initial porosities. \citet{2016arXiv160305979L} found that if a planetesimal forms too early in the age of the solar system, the high initial $^{26}$Al content will cause the body to melt and differentiate. Therefore we also assume that our fiducial model planetesimal is not early forming (i.e. t$_{form} >$ 1.4 Mya).

The initial concentrations of reactants for our chemical simulations were chosen to match the initial concentrations in a previous study of biomolecule synthesis within meteorite parent bodies \citep{Reference44}, which are based on the mixing ratios (mol X/mol H$_2$O) spectroscopically measured in comets \citep{2011ARA&A..49..471M,2004come.book..391B,2004A&A...418.1141C,2000ARA&A..38..427E,2000A&A...353.1101B}. For molecules not available in \citet{Reference44}, initial concentrations were taken from the molecular abundances spectroscopically measured in comet Hale-Bopp. When two concentrations were provided, the average was taken between the two. For H$_2$, where its presence in comets is thought to be significant, but is only proven from the identification of rotationally resolved molecular hydrogen transitions \citep{Reference21}, a value matching the abundance of CO is chosen---but is adjusted during the experiment to see how strong its variation can affect production. (CO is a commonly used as a tracer of H$_2$ in interstellar clouds, as they are the two most abundant molecules in such environments.)

The initial concentrations for all the simulation reactants are shown in Table~\ref{InitialConcentrations}.

\begin{table}[!ht]
\centering
\caption{Initial concentrations of the reactants in our model planetesimal. 
Molecular abundances are in percent normalized to water and, when possible, 
are made to match the initial concentrations from previous biomolecule simulations for meteorite parent body environments \citep{Reference44}. For 
initial concentrations not used in \citet{Reference44}, concentrations, with the exception of H$_2$, are based on molecular abundances measured in comet Hale-Bopp \citep{2011ARA&A..49..471M,2004come.book..391B,2004A&A...418.1141C,2000ARA&A..38..427E,2000A&A...353.1101B}. The initial concentration of H$_2$ is taken to match that of CO and will 
be adjusted to check for sensitivity.\label{InitialConcentrations}}
\begin{tabular}{lc}
\\
\multicolumn{1}{l}{Molecule} &
\multicolumn{1}{c}{Concentration (mol X/mol H$_2$O)}\\ \hline
H$_2$O & 100 \\
CO & 17.5 \\
H$_2$ & 17.5\\
NH$_3$ & 0.7\\
HCN & 0.25\\
Formic Acid & 0.075\\
CH$_2$O & 0.066\\
Formamide & 0.015\\
\end{tabular}
\end{table}

\section{Results}\label{resultssec}

Figures~\ref{AdenineSim}, \ref{CytUraSim}, and \ref{GuaThySim} display the nucleobase abundances from the simulations of A, C, U, G, and T synthesis. Additional simulation models are added to these figures with reduced reactant concentrations (one-twenty thousandth) to the fiducial values in Table~\ref{InitialConcentrations}. Since all initial concentrations are given as a percent fraction with respect to water, when a simulation is run with a reduced water concentration, all initial concentrations in the simulation are reduced by the same fraction.

One NC nucleobase reaction is chosen for each nucleobase to be simulated while also allowing the Strecker synthesis of glycine \citep{Reference44} to occur. In these simulations, the NC nucleobase reactions compete for reactants with the formation of glycine, which has some similar reactants (H$_2$O, HCN, CH$_2$O, NH$_3$). Glycine is specifically chosen instead of another amino acid because it forms from formaldehyde, which is a reactant in NC nucleobase synthesis. Glycine is also the most abundant in meteorites and thus represents one of the more difficult opponents for a competition simulation.

Figure~\ref{LaRoweSim} below displays an additional competition simulation, where all five NC nucleobase reactions from a theoretical nucleobase synthesis study \citep{2008OLEB...38..383L} plus an additional NC thymine reaction \citep{Reference53} are allowed to synthesize. The intention is to see how the six NC nucleobase reactions would compete with each other for similar reactants. A competition simulation is also run for the three FT reactions \citep{1968GeCoA..32..175H}, though this model very poorly represents the relative experimental yields of FT synthesis and cannot be used for insightful analysis.

\subsection{Adenine}\label{adenine}

\begin{figure*}[hbtp]
\centering
\includegraphics[width=\textwidth]{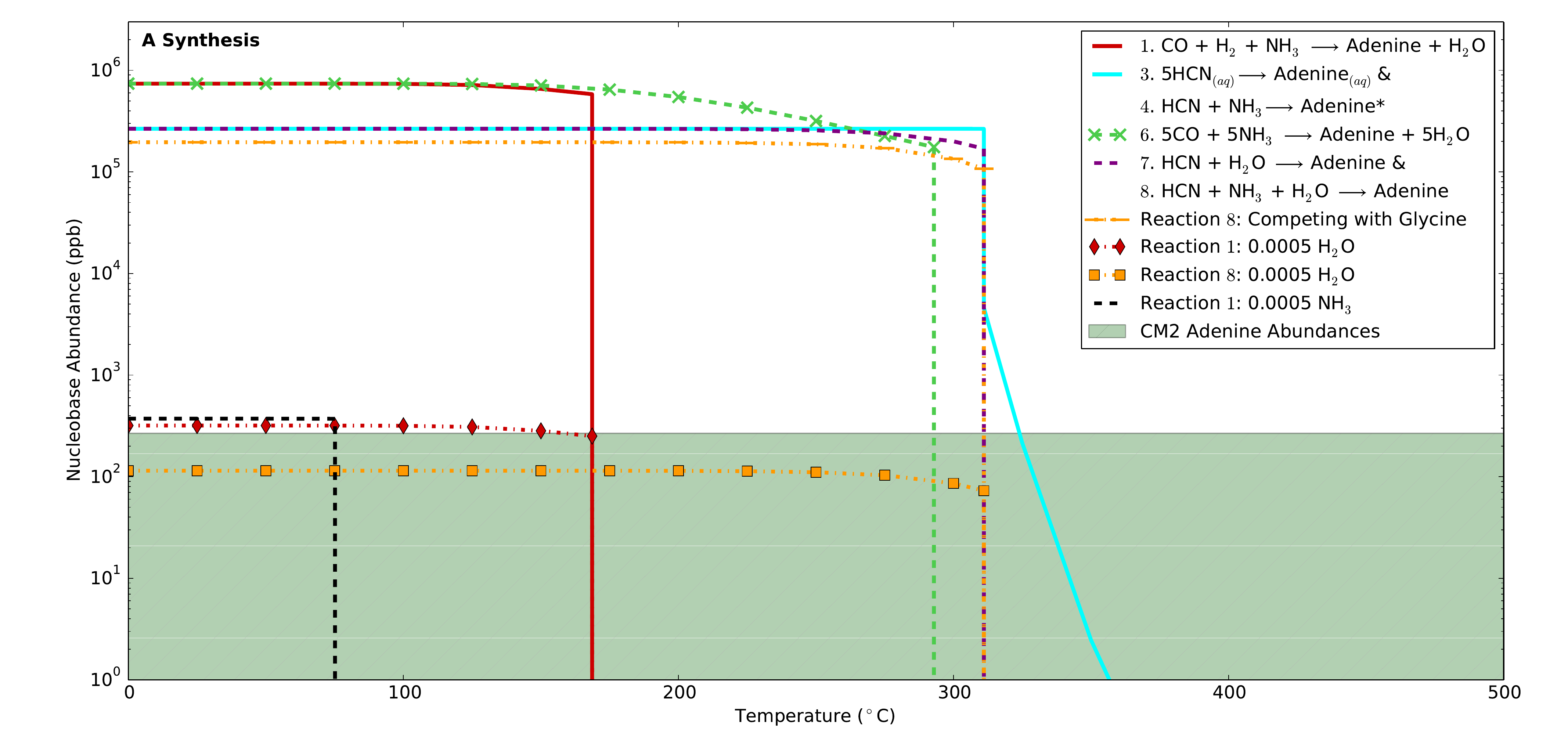}
\caption{Theoretical adenine abundances from simulations of the adenine candidate reactions in Table~\ref{Candidates}---except for reaction 24 which is unproductive at all temperatures. Reactions 3 and 4, and reactions 7 and 8 have equivalent curves. Additional models for reactions 1 and 8 are simulated with reduced reactant concentrations to the fiducial values in Table~\ref{ReactionRates}. One model is included which allows the Strecker synthesis of glycine \citep{Reference44} to react alongside reaction 8. All simulations were run at 100 bar, in intervals of 25$^{\circ}$C from 0--500$^{\circ}$C. The shaded box represents the range of adenine abundances in CM2 meteorites \citep{2011PNAS..10813995C,1981GeCoA..45..563S}. *Reaction 4 was simulated as an aqueous reaction---even though experiments performed these reactions without water \citep{Reference41,ReferenceWaka}---due to the lack of non-aqueous Gibbs free energy data for HCN.}
\label{AdenineSim}
\end{figure*}

In Figure~\ref{AdenineSim} we display the individual simulation results from the six productive A reaction pathways in Table~\ref{Candidates}. Reaction 1, the FT synthesis of A, and reaction 6, the NC synthesis of A, are the most productive reactions of the seven, with resultant abundances just over 7x10$^5$ ppb before 165$^{\circ}$C and 290$^{\circ}$C respectively. These are both gas phase reactions with CO and NH$_3$ as reactants, but reaction 1 also requires a catalyst and the reactant H$_2$. The similarity in the A abundances from these two reactions may suggest that H$_2$ is unnecessary for producing A within planetesimals. On the other hand, the lowest temperature in which A was synthesized in the lab using reaction 6 was 500$^{\circ}$C \citep{1968GeCoA..32..175H}---a temperature for which A decomposes in $<$ 1 second (see Table~\ref{ReactionRates}). Perhaps the catalyst in reaction 1 is necessary to produce A from CO and NH$_3$ at temperatures less than 500$^{\circ}$C.

Reactions 3, 4, 7 and 8 all produce near 3x10$^5$ ppb of A across all temperatures less than 300$^{\circ}$C. Since the only difference between reactions 4 and 3 (as well as reactions 8 and 7) is the inclusion of NH$_3$ as a reactant, the equivalent A-production curves for these reactions hints at the unimportance of NH$_3$ in HCN-based reactions at equilibrium. This result is consistent with laboratory results, which produce similar max yields of A with (0.05$\%$) \citep{Reference27} and without (0.04$\%$) \citep{Reference12} NH$_3$ as a reactant. The FT reaction for A is approximately three times more productive than these NC reactions. Every aqueous NC reaction becomes unproductive at the liquid-to-gas phase transition of water. Reaction 24 (the CA reaction) is not present in Figure~\ref{AdenineSim}, as it is unproductive.

It is quite noticeable that the individual reaction simulations produce much higher abundances of A than is measured in carbonaceous chondrites. The abundances of A measured in CM2 meteorites \citep{2011PNAS..10813995C,1981GeCoA..45..563S} (green shaded region) are at least 3 orders of magnitude lower than the least productive A reaction. In order to try to account for this over-production, both reactions 1 and 8 are modeled with one-two thousandth the fiducial initial concentration of water. This reduction in water causes reaction 8 to produce an amount of A within the meteoritic abundance range, and reaction 1 to produce an amount near the boundary. 

Since the fiducial concentration of H$_2$ is chosen to match that of its tracer molecule (CO), we adjust the H$_2$ concentration for reaction 1 by various amounts to see how it affects the amount of A produced. The results are not shown in Figure~\ref{AdenineSim} because adjusting the concentration of H$_2$ by orders of magnitude in either direction is found to have a very minor effect in the production of A. This is because H$_2$ is not the limiting reagent of FT synthesis, and thus doesn't have as large of an effect on nucleobase production as does NH$_3$. Adjusting the amount of NH$_3$ by one-two thousandth the fiducial initial concentration (see black dotted line in Figure~\ref{AdenineSim}) leads to an almost equivalent decrease in the production of A as does one-two thousandth the water (and hence one-two thousandth the scaling reactants). The same NH$_3$ adjustment was made for simulations of the FT synthesis of G and C, verifying NH$_3$ as the limiting reagent for all the FT reactions in this study.

Simulation results from allowing the Strecker synthesis of glycine to compete against reaction 8 are also shown in Figure~\ref{AdenineSim}. This leads to only a slight decrease in the production of A (by about a factor of 1.4).

\subsection{Cytosine and Uracil}

\begin{figure*}[hbtp]
\centering
\includegraphics[width=\textwidth]{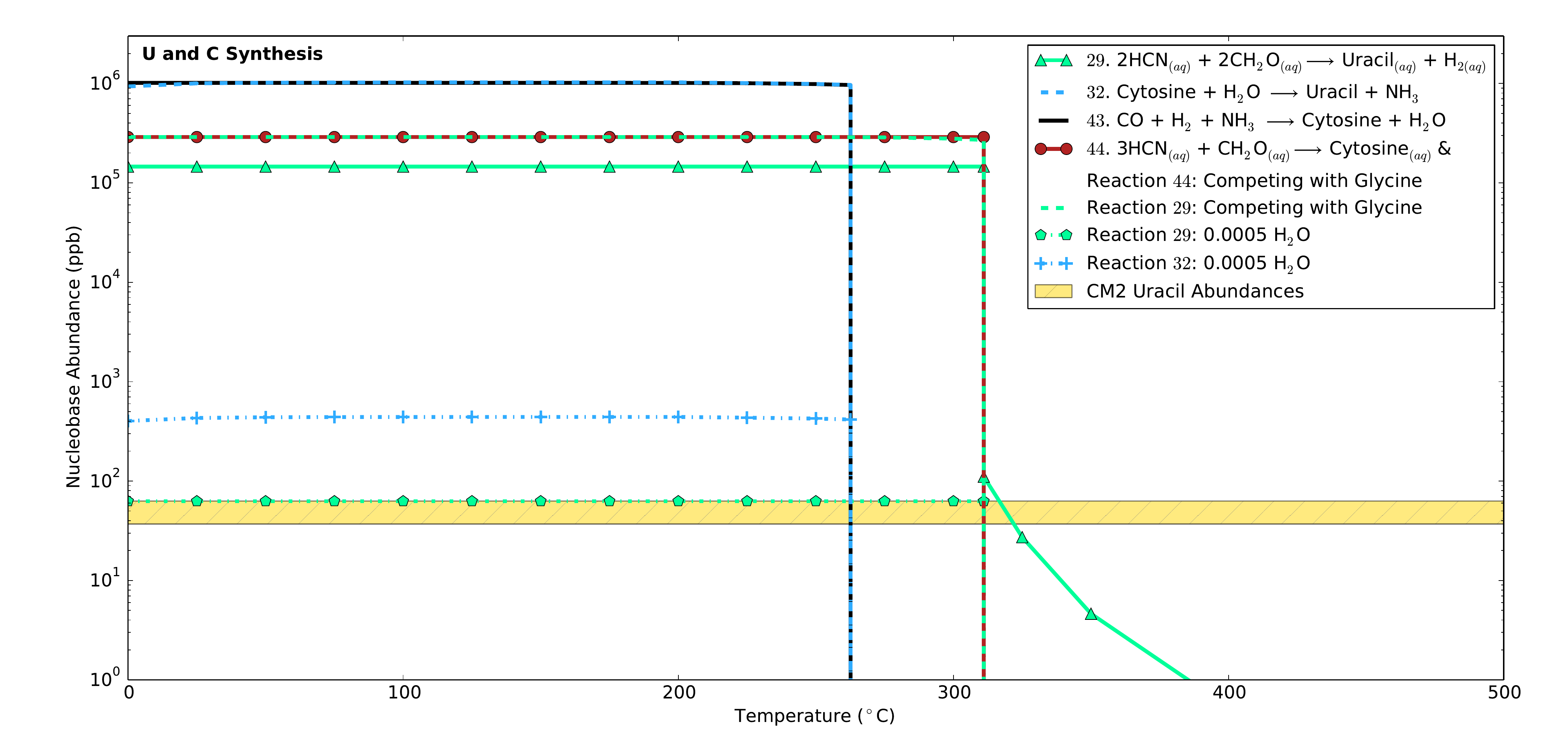}
\caption{Theoretical cytosine and uracil abundances from simulations of their candidate reactions in Table~\ref{Candidates}---except for reaction 49 which is unproductive at all temperatures. Additional models for reactions 29 and 32 are simulated with reduced water (and scaling reactant) concentrations to the fiducial values in Table~\ref{InitialConcentrations}. Two models are included which allow the Strecker synthesis of glycine \citep{Reference44} to react alongside each nucleobase's candidate NC reaction (reaction 44 for cytosine, reaction 29 for uracil). Reaction 44 and reaction 44 in competition with the Strecker synthesis of glycine have equivalent curves. All simulations were run at 100 bar, in intervals of 25$^{\circ}$C from 0--500$^{\circ}$C. The shaded box represents the range of uracil abundances in CM2 meteorites \citep{1979Natur.282..709S}.}
\label{CytUraSim}
\end{figure*}

In Figure~\ref{CytUraSim} we display the nucleobase abundances from the individual simulations of C and U synthesis. Reaction 43 (FT synthesis) is the most productive C reaction, producing abundances of 10$^6$ ppb at temperatures below $\sim$260$^{\circ}$C. Reaction 44 (NC synthesis) produces approximately a factor of 3 less C than reaction 43, and synthesizes up to the boiling point of water at 100 bar (at 311.03$^{\circ}$C). Since the temperatures where the production of C ceases for reactions 43 and 44 are both above the approximate upper boundary of nucleobase synthesis within planetesimals (165$^{\circ}$C, see Section~\ref{stability}), the difference in the temperature where nucleobase synthesis shuts off between these reactions is likely insignificant. Reaction 48 (the CA reaction) is not present in Figure~\ref{CytUraSim}, as it is unproductive at all temperatures.

The input C concentration for the simulation of reaction 32 (deamination of C) is set as the output abundance of C from the simulation of reaction 43 (FT Synthesis of C). The purpose is to see the percentage of C produced in a planetesimal that would decompose into U once the planetesimal reaches equilibrium. This leads to a significant result: the curves for reaction 32 and 43 are nearly identical, meaning nearly all of the C in a planetesimal deaminates into U once equilibrium is reached. This is significant because C is not found within meteorites, and here we can see that all of the cytosine decomposes into uracil within a model meteorite parent body at chemical equilibrium. 

Since the deamination of C simulation produces an amount of U essentially equivalent to the reactant C concentration, and both reactions for C are more productive than reaction 29 for U, this makes the deamination of C into U the most productive U reaction. Reaction 29 of U produces approximately 1.5x10$^5$ ppb, which is about an order of magnitude less than reaction 32 of U.

The two U reactions are also simulated with one-two thousandth the fiducial concentration of water, and are illustrated in Figure~\ref{CytUraSim}. Reaction 29 with a reduced water concentration fits on the borderline of the CM2 meteoritic U abundances \citep{1979Natur.282..709S} (yellow shaded region). Reaction 32 modelled with one-two thousandth the initial water produces about an order of magnitude more U than the meteoritic abundance.

Finally, the competition reactions allowing the Strecker synthesis of glycine to occur alongside the NC synthesis of C and U are also shown in Figure~\ref{CytUraSim}. Reaction 44 appears to produce the same amount of C regardless of whether glycine is also a permitted product. Interestingly, reaction 29 produces twice as much U when competing with glycine as it did in its individual reaction. Further analysis reveals that more favourable pathways for U synthesis open up when glycine is also allowed to synthesize, allowing the secondary product in reaction 29, H$_2$, to be exploited in producing glycine and additional U.

\subsection{Guanine and Thymine}

\begin{figure*}[hbtp]
\centering
\includegraphics[width=\textwidth]{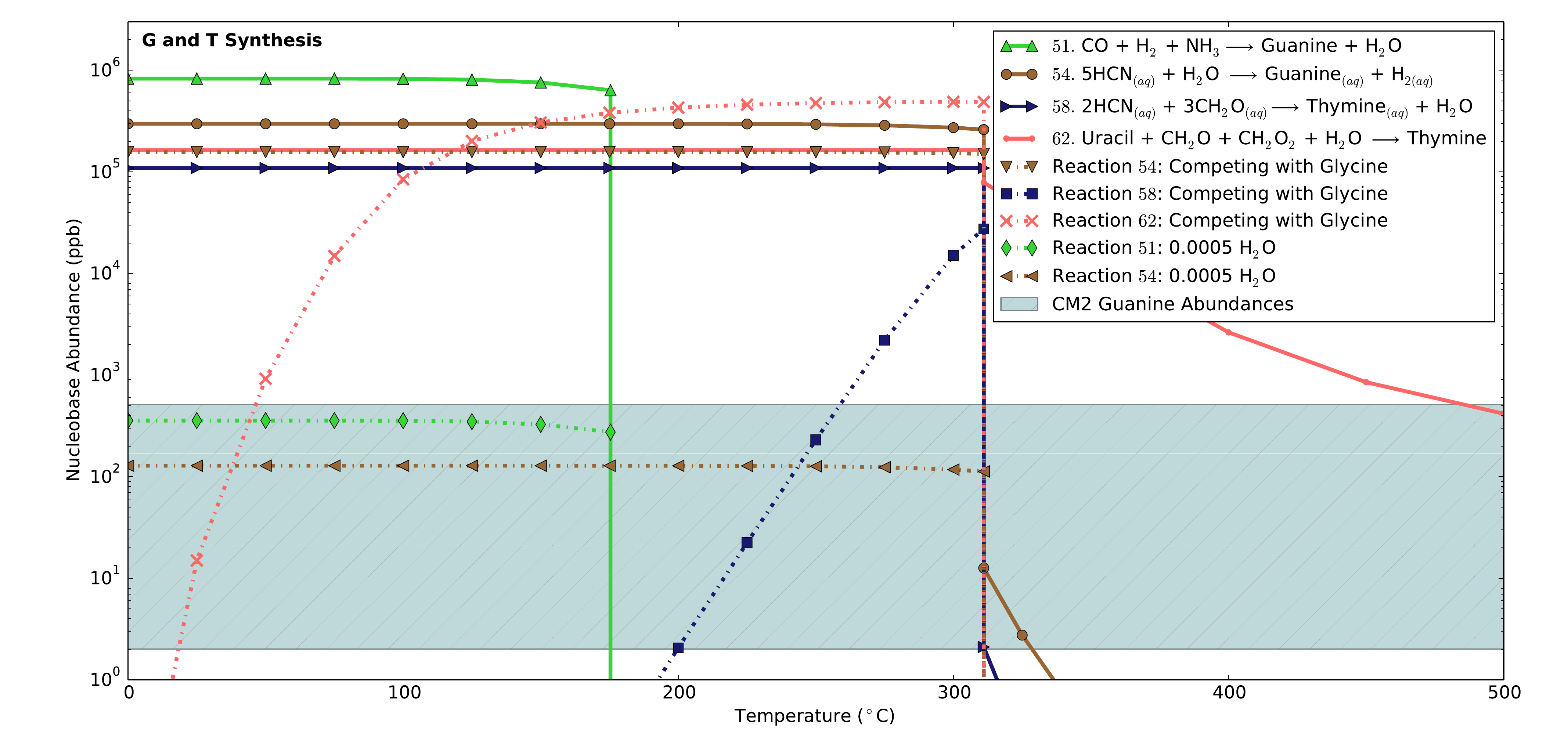}
\caption{Theoretical guanine and thymine abundances from simulations of their candidate reactions in Table~\ref{Candidates}. Additional models for reactions 51 and 54 are simulated with reduced water (and scaling reactant) concentrations to the fiducial values in Table~\ref{InitialConcentrations}. Three models are included which allow the Strecker synthesis of glycine \citep{Reference44} to react alongside each nucleobase's candidate NC reaction (reaction 54 for guanine, reactions 58 and 62 for thymine). All simulations were run at 100 bar, in intervals of 25$^{\circ}$C from 0--500$^{\circ}$C. The shaded box represents the range of guanine abundances in CM2 meteorites \citep{2011PNAS..10813995C,Reference37,1981GeCoA..45..563S,1977GeCoA..41..961V}.}
\label{GuaThySim}
\end{figure*}

In Figure~\ref{GuaThySim} we display the individual simulation results for the synthesis of G and T.  Reaction 51 (FT synthesis) is found to be the most productive reaction for G, with abundances near 8x10$^5$ ppb at temperatures less than $\sim$175$^{\circ}$C. Reaction 54 (NC synthesis) is about a factor of 3 less productive than reaction 51, producing up to the liquid-to-gas phase transition of water at 100 bar (311.03$^{\circ}$C). Since both 175$^{\circ}$C and 311.03$^{\circ}$C are above the approximate 165$^{\circ}$C upper boundary of nucleobase synthesis within planetesimals (as estimated in Section~\ref{stability} above), the difference between these temperatures where G synthesis shuts off is probably insignificant.

The NC candidate T reactions (nos. 58 and 62), produce approximately 1--2x10$^5$ ppb when simulated individually. These reactions, along with reaction 29 of U, are the least productive individual simulations.

Reactions 51 and 54 of G are modelled with one-two thousandth the fiducial water concentration.  Both reactions with reduced water fall into the CM2 meteoritic abundance range for G \citep{2011PNAS..10813995C,Reference37,1981GeCoA..45..563S,1977GeCoA..41..961V} (blue shaded region).

The NC reactions for G and T are also separately simulated in competition with the Strecker synthesis of glycine and their abundances are illustrated in Figure~\ref{GuaThySim}. Reaction 54 when competing with glycine produces about a factor of 2 less G than did the individual reaction 54 simulation. This is not a substantial reduction when abundances are in the 10$^5$ ppb range, but it is still worth noting that molecular competition for reactants could be a contributor to the decrease in G production within planetesimals.

Much more significant is the effect of molecular competition on reaction 58 of T. Allowing the glycine Strecker reaction to share reactants with reaction 58 makes the latter completely unproductive at temperatures lower than 200$^{\circ}$C. Because 200$^{\circ}$C is above the approximate 165$^{\circ}$C upper boundary of nucleobase synthesis within planetesimals, as estimated in Section~\ref{stability}, the competition for reactants with glycine makes reaction 58 of T theoretically unproductive within planetesimals.

Reaction 62 of T on the other hand is only greatly affected by the molecular competition with glycine at the lowest temperatures in this simulation ($\sim$0--25$^{\circ}$C). This means that  there should still be one productive reaction pathway for T within planetesimals in spite of molecular competition with the Strecker synthesis of glycine.

\subsection{Nucleobase Reactions Simulated Together}

\begin{figure*}[hbtp]
\centering
\includegraphics[width=\textwidth]{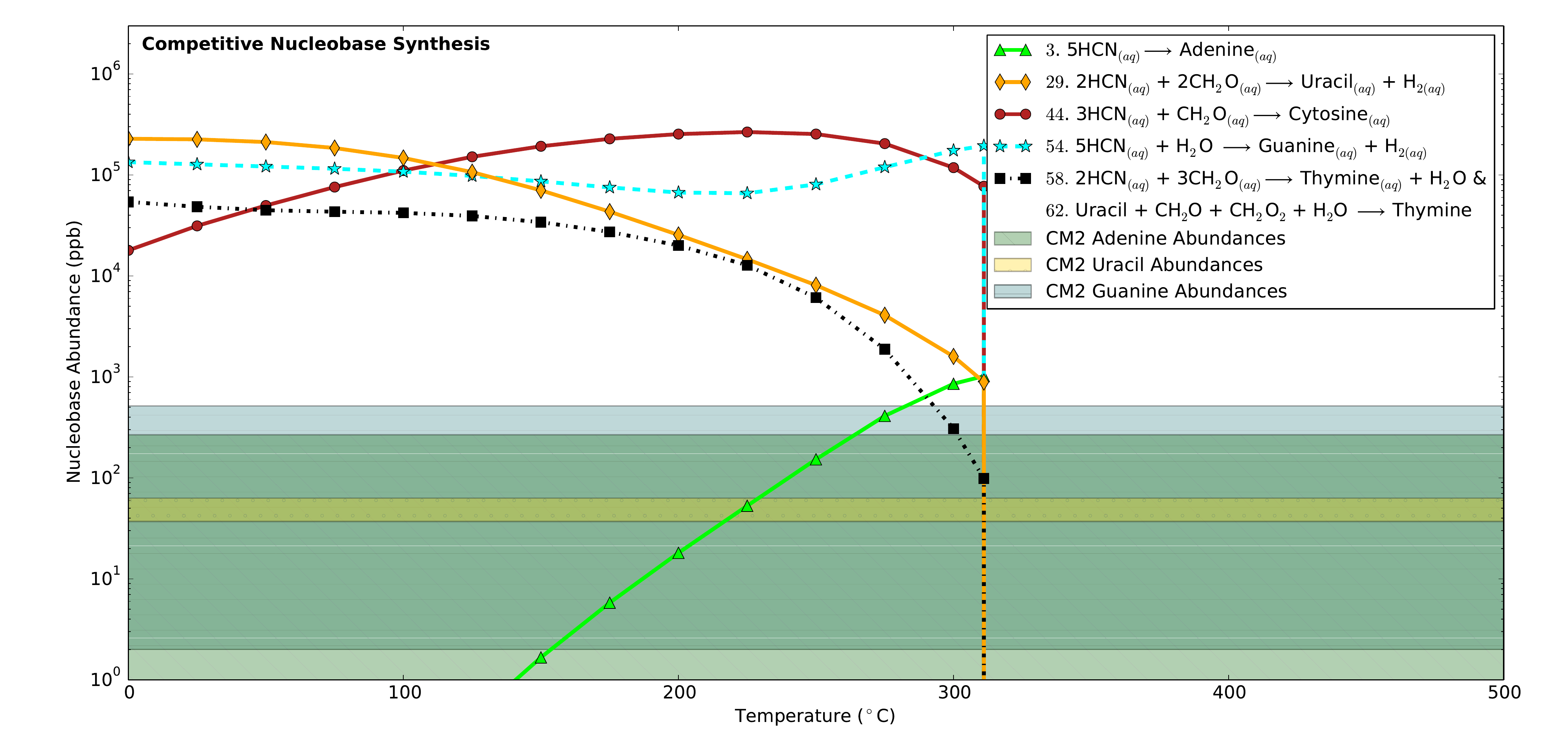}
\caption{Theoretical guanine, adenine, uracil, cytosine and thymine abundances from the competition simulation including each of the NC nucleobase reactions from a theoretical study \citep{2008OLEB...38..383L} plus an additional NC thymine reaction \citep{Reference53} (reactions 3, 29, 44, 54, 58 and 62 in Table~\ref{Candidates}). The black dotted curve represents the thymine produced from both reactions 58 and 62. All simulations were run at 100 bar, in intervals of 25$^{\circ}$C from 0--500$^{\circ}$C. The shaded boxes represent the range of guanine, adenine and uracil abundances in CM2 meteorites.}
\label{LaRoweSim}
\end{figure*}

In Figure~\ref{LaRoweSim} we illustrate how nucleobases might compete with each other for reactants. The results are from a single simulation of each of the the five proposed NC nucleobase reactions from a theoretical study \citep{2008OLEB...38..383L} and an additional NC thymine reaction \citep{Reference53}. Reaction 29 of U is the most productive reaction in this competition simulation until just after 100$^{\circ}$C, producing approximately 2x10$^5$ ppb. As temperatures increase from there, reaction 44 of C becomes the most productive reaction producing $\sim$2--3x10$^5$ ppb. Since we already know from Figure~\ref{CytUraSim} that practically all of the C produced in planetesimals would deaminate into U at equilibrium, the C curve should be added to the U curve to get the complete U abundance in this competition reaction. This makes U the most productive nucleobase at all temperatures when competing against the other nucleobases for reactants.

The second most abundant nucleobase when all NC nucleobase reactions are run together is G, with abundances near 10$^5$ ppb. Then for temperatures less than 200$^{\circ}$C, T is the third most abundant nucleobase with its two reactions producing a combined abundance of $\sim$2--5x10$^4$ ppb. Lastly, reaction 3 of A is the least productive NC reaction in this competition simulation, only producing after 150$^{\circ}$C in abundances of 1--1000 ppb.
The production of A actually fits into the range of its meteoritic abundances just by competing with the other four nucleobases. This makes the reduced water models for A synthesis perhaps unnecessary in explaining the superfluous abundances of A (with respect to the meteoritic record) from individual reaction simulations.

Finally, the competition reaction of the three FT nucleobase reactions \citep{1968GeCoA..32..175H} were run in a single simulation. Unfortunately the results showed that no G or A would be produced at any temperature in the 0--500$^{\circ}$C range, and that C would be the only nucleobase produced. This does not conform with the laboratory results \citep{1968GeCoA..32..175H} or the meteoritic record \citep{2011PNAS..10813995C,Reference37,1981GeCoA..45..563S,1977GeCoA..41..961V}. This competition simulation is therefore not a good model for FT synthesis, and likely requires the inclusion of one or several of the additional molecules produced from these laboratory experiments (e.g. urea, melamine, guanidine) to lower the thermodynamic favourability of C so that the other nucleobases can also produce. For further discussion regarding the caveats of modeling competition between reactions, see Appendix B.

\subsection{Relative Nucleobase Abundances}\label{accuracy}

\begin{figure}[ht!]
\centering
\includegraphics[width=80mm]{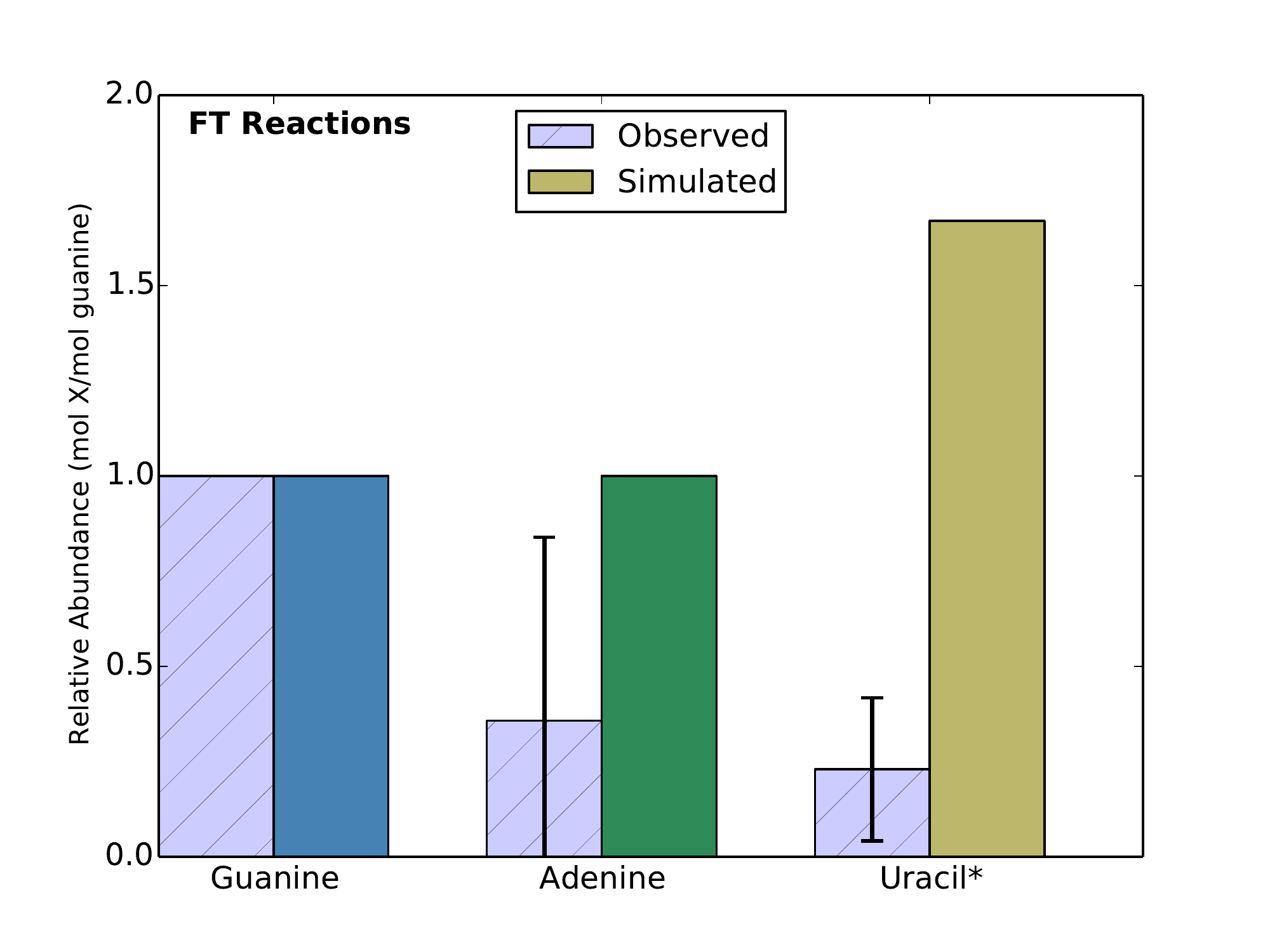}
\caption{Relative nucleobase abundances of guanine, adenine and uracil from CM2 meteorites and individual Fischer-Tropsch (FT) simulations. Observed relative nucleobase abundances in CM2 meteorites are in light blue with black horizontal stripes and black error bars. Simulation abundances for FT reactions at 100$^{\circ}$C and 100 bar are in blue (guanine), green (adenine) and gold (uracil). *Uracil simulation abundance is from the FT cytosine reaction decomposing into uracil (reaction 32).}
\label{RelativeFT}
\end{figure}

\begin{figure}[ht!]
\centering
\includegraphics[width=80mm]{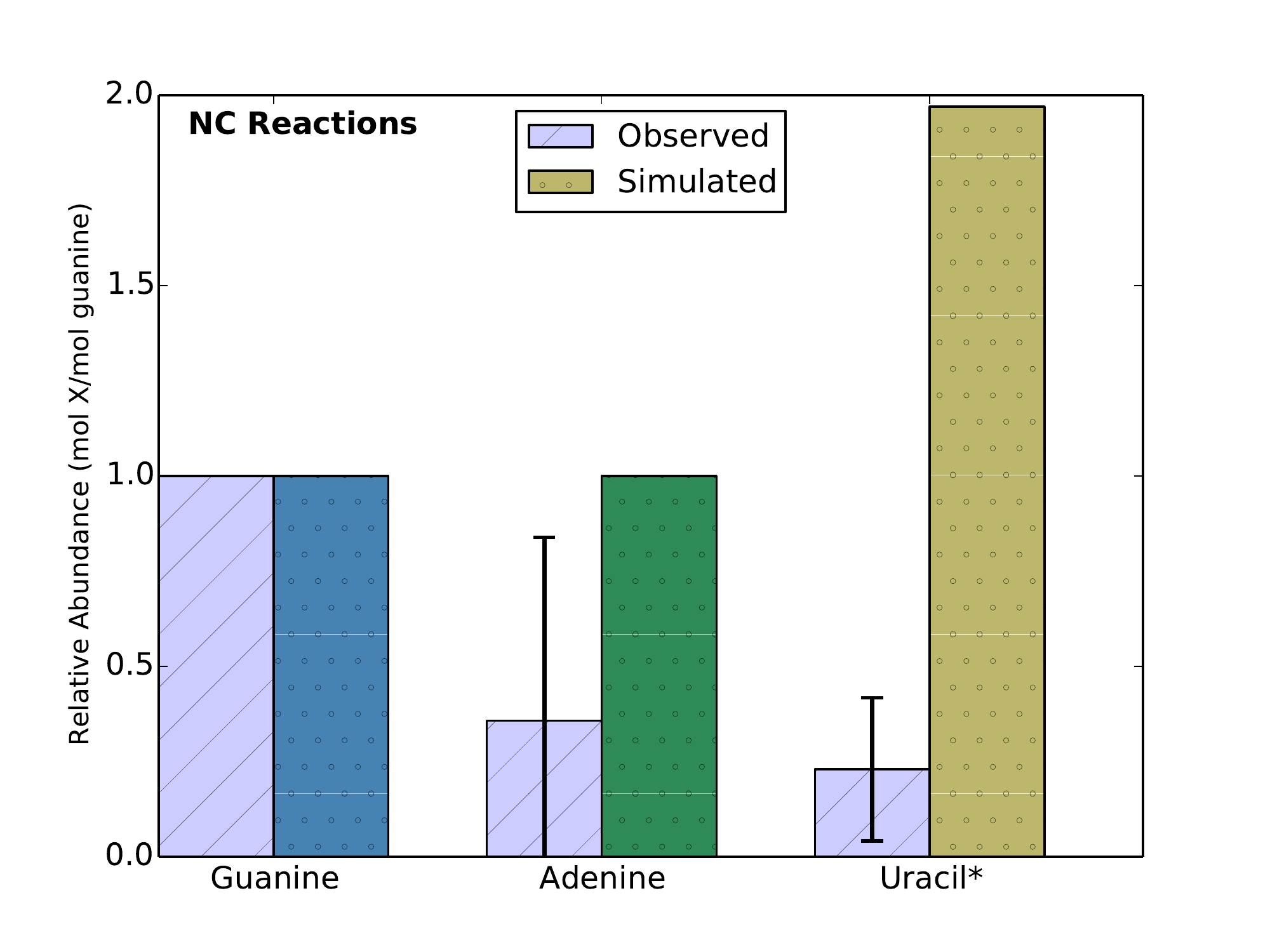}
\caption{Relative nucleobase abundances of guanine, adenine and uracil from CM2 meteorites and individual non-catalytic (NC) simulations. Observed relative nucleobase abundances in CM2 meteorites are in light blue with black horizontal stripes and black error bars. Simulation abundances for NC reactions at 100$^{\circ}$C and 100 bar are in blue (guanine), green (adenine) and gold (uracil) with black dots. *Uracil simulation abundance is from the NC uracil simulation (reaction 29) plus the NC cytosine reaction decomposing into uracil (reaction 32).}
\label{RelativeNC}
\end{figure}

In Figures~\ref{RelativeFT} and \ref{RelativeNC}, the relative nucleobase abundances from individual FT and NC synthesis simulations are compared with the relative nucleobase abundances in CM2 meteorites \citep{Reference46}. This comparison helps us determine how well our nucleobase synthesis simulations conform to the meteoritic record. Relative abundances are in moles of nucleobase over moles of guanine. Relative simulation abundances are calculated at 100 $^{\circ}$C and 100 bar.

In Figure~\ref{RelativeFT}, the relative A to G abundance from our individual FT synthesis simulations is exactly 1. This value is slightly outside of the error of the relative A to G within CM2 meteorites of 0.36 $\pm$ 0.48 \citep{Reference46}. There is no FT synthesis simulation for U, which has meteoritic abundances, but there is a FT synthesis simulation for C. Since our simulation of reaction 32 has shown that C completely decomposes into U in aqueous solution at equilibrium (Figure~\ref{CytUraSim}), we estimate the relative U to G abundance for FT synthesis as the relative C to G abundance for FT synthesis. Using this method, the relative U to G abundance for FT synthesis is 1.67. This value is several sigma outside of the error bars of the relative U to G within CM2 meteorites of 0.23 $\pm$ 0.19 \citep{Reference46}. This discrepancy is discussed in Section~\ref{discrepancy}.

%We also notice a discrepancy when comparing the these relative simulation abundance ratios with the relative nucleobase yields from the laboratory experiments \citep{1968GeCoA..32..175H}. Relative laboratory yields which are 1.78 for the relative A to G and---if replacing the C value for the U value to represent C completely deaminating into U within a planetesimal---0.56 for the relative U to G.

In Figure~\ref{RelativeNC} we see that the relative A to G abundance from our individual NC synthesis simulations is 1. For consistency with the relative U to G abundance calculation for FT synthesis, the relative U to G abundance for NC synthesis is calculated as the sum of the NC U and C reaction abundances, divided by the NC G reaction abundance. This again is because C has been shown to completely decompose into U at equilibrium due to hydrolysis. Using this method, the relative U to G abundance for NC synthesis is 1.97. These relative abundances are quite similar to the relative abundances produced via FT reactions.

When the NC synthesis of A, G, C, U and T (both reactions 58 and 62) are allowed to occur together in the same simulation, the relative U to G abundance is 1.85. This again is very similar to the relative U to G abundances from individual FT and NC nucleobase simulations. On the other hand, the relative A to G abundance from this competition simulation drops to 0. This satisfies the lower errorbar of the relative A to G in CM2 meteorites of 0.36 $\pm$ 0.48 \citep{Reference46}, though a complete lack of A is rare within carbonaceous chondrite nucleobase assays. Because of the lack of A production in this competition simulation, we suggest that individual nucleobase synthesis (i.e. weak coupling) simulations probably provide more accurate results than simulating nucleobases in competition.

\section{Discussion}\label{discussionsec}

The principle result obtained from these simulations is that nearly all of the C produced within our model planetesimal deaminates into U (reaction 32) once equilibrium is reached. This reaction occurs relatively quickly, decomposing half of the C into U in 3.5 hours at 165$^{\circ}$C, or 17,000 years at 0$^{\circ}$C \citep{Reference45}. Since planetesimals are thought to have had aqueous interiors for millions of years \citep{2005E&PSL.240..234T}, and C decomposes in an aqueous environment in less than 17,000 years, C should effectively never be found in carbonaceous chondrites. 

This result demands an explanation for the emergence of an RNA World which doesn't involve the meteoritic delivery of C. One possibility could be that the first self-replicating molecules on Earth formed from nucleotides that obtained C from somewhere other than meteorites or comets. For example, C has been synthesized in the laboratory by exposing icy interplanetary dust analogs containing pyrimidine to UV radiation under space-like conditions \citep[and references therein]{2014ApJ...793..125N}. It has also been estimated that interplanetary dust particles could have had influxes in the range of $\sim$10$^8$ kg yr$^{-1}$ at the time of the origins of life \citep{1992Natur.355..125C}. Although icy dust particles present a potential source of prebiotic C, it must be noted that no nucleobases have yet been detected on the surfaces of ices in space. Just recently, the gas chromatograph and time-of-flight mass spectrometer aboard Rosetta's Philae lander ``sniffed'' for organic compounds on the surface of comet 67P/Churyumov-Gerasimenko \citep{2015Sci...349b0689G}. Of the 16 organic compounds detected, none were nucleobases.

Alternatively, the first RNA molecules could have formed from nucleotides that synthesized without the use of nucleobases (e.g. from cyanamide, cyanoacetylene, glycolaldehyde, glyceraldehyde, inorganic phosphate and UV radiation \citep{Reference56}). It is also possible that the first RNA molecules didn't involve C at all. {\it In vitro} evolution has been used in the lab to obtain catalytic RNA molecules (ribozymes) that only contain adenosine, guanosine, and uridine nucleotides \citep{Reference65} and only uridine and 2,6-diaminopurine nucleotides \citep{Reference66}.

Another significant result from these simulations is that the NC synthesis of T from U, formaldehyde and formic acid (reaction no. 62) is thermodynamically favorable for planetesimal-like conditions. This reaction has produced $\sim$10$^5$ ppb of T when simulated individually, and when in competition with reactions that have similar reactants---such as the Strecker synthesis of glycine or the five other NC nucleobase reactions. The other two candidate T reactions (nos. 58 and 63) were either completely unproductive, or unproductive until 200$^{\circ}$C when competing with the Strecker synthesis of glycine.

\subsection{Thermodynamic Favourability of Thymine Synthesis}

T is not found in meteorites, thus it is curious to see a favourable reaction pathway for T in both our individual and competition simulation results. But is thermodynamic favourability enough to justify T production within planetesimals? In Table~\ref{Balanced} we take a closer look at the $\Delta G_r$ of reaction 62 of thymine (T) in comparison to that of each reaction type for adenine (A) and guanine (G).

To obtain the most valid $\Delta G_r$ for each reaction, it is important that the reaction equations balance. A balanced equation means that the number of each atom on the left side of the equation equals the number of each atom on the right side. Theoretical balanced equations for reaction 62 of T, the FT, NC and CA synthesis of A, and the FT and NC synthesis of G are shown in Table~\ref{Balanced}. Reaction equations for A and G balance fairly simply using only their reactant and product molecules from Table~\ref{Candidates}, but reaction 62 of T requires an additional product to balance. We chose CO$_2$ as this product because the authors of the corresponding paper \citep{Reference53} suggested decarbonation to be an intermediate step of this reaction.

\begin{table*}[t]
\centering
\caption{Balanced reaction pathways for reaction 62 of thymine (NC), the FT, NC and CA syntheses of adenine, and the FT and NC syntheses of guanine along with their Gibbs free energies of reaction $\Delta G_r$ at 50$^{\circ}$C and 100 bar.\label{Balanced}}
\begin{tabular}{llc}
\\
\multicolumn{1}{l}{Type} &
\multicolumn{1}{l}{Balanced Reaction} &
\multicolumn{1}{c}{$\Delta G_r$ (kJ/mol)}\\ \hline
$\underline{Adenine}$ & \\
FT & 5CO + 5NH$_{3}$ $\xrightarrow{NiFe+||Al_{2}O_{3}+||SiO_{2}}$ C$_{5}$H$_{5}$N$_{5}$ + 5H$_{2}$O & -158\\
NC & 5HCN$_{(aq)}$ $\rightarrow$ C$_5$H$_5$N$_{5(aq)}$ & -276\\
CA & 5CH$_3$NO $\xrightarrow{Al_{2}O_{3}||SiO_{2}}$ C$_5$H$_5$N$_5$ + 5H$_{2}$O & 102\\
$\underline{Guanine}$ \\
FT & 5CO + 5NH$_{3}$ $\xrightarrow{NiFe+||Al_{2}O_{3}+||SiO_{2}}$ C$_5$H$_5$N$_5$O + 4H$_{2}$O + H$_{2}$ & -194\\
NC & 5HCN$_{(aq)}$ + H$_{2}$O $\rightarrow$ C$_5$H$_5$N$_5$O$_{(aq)}$ + H$_{2(aq)}$ & -264\\
$\underline{Thymine}$ \\[+1mm]
NC & C$_{4}$H$_{4}$N$_{2}$O$_{2}$ + CH$_{2}$O$_{2}$ + CH$_{2}$O $\rightarrow$ C$_{5}$H$_{6}$N$_{2}$O$_{2}$ + CO$_{2}$ + H$_{2}$O & -146\\
\hline
\multicolumn{3}{l}{\footnotesize NC: Non-catalytic, CA: Catalytic, FT: Fischer-Tropsch.}
\end{tabular}
\end{table*}

Both the NC syntheses of G and A are more energetically favourable than the FT syntheses of G and A, as made apparent by the former's lower $\Delta G_r$. This means that FT synthesis simulations likely did not produce more G and A than NC synthesis simulations because of energetic considerations. Also notice how the CA synthesis of A requires input energy to occur (102 kJ/mol), thus being unproductive at equilibrium. This calculation conforms to our unproductive CA simulations of A synthesis.

%The NC synthesis of A is $\sim$1.7 times more favourable than the FT synthesis of A, and the NC synthesis of G is $\sim$1.4 times more favourable than the FT synthesis of G.

Reaction 62 of T is relatively close in $\Delta G_r$ to the FT synthesis of A, meaning that neither is really favoured over the other from a thermodynamic standpoint. Yet 4.5 times more A is produced from FT synthesis than T is produced from reaction 62. This again illustrates that thermodynamic favourability is not the only factor when considering how productive a reaction will be at equilibrium. For this reason, we look closer at input reactant concentrations to see how important they are in determining equilibrium nucleobase abundances.

\subsection{Effects of Initial Reactant Concentrations}

To consider how input reactant abundances effect nucleobase production at equilibrium, we compare how simulation nucleobase abundance ratios relate to initial limiting reagent concentration ratios for various reactions. The limiting reagent of a reaction is the reactant that is completely used up at equilibrium, and therefore limits how much of the nucleobase can be produced. Table~\ref{LimitingReagent} lists these ratios for the FT to NC reactions of A and G, and the FT synthesis of A to reaction 62 of T. The limiting reagents for FT synthesis, NC synthesis, and reaction 62 of T are NH$_3$, HCN, and H$_2$CO, respectively.

\begin{table*}[t]
\centering
\caption{This is a comparison table between simulation abundance ratios and initial limiting reagent concentration ratios for various reactions. Nucleobase simulation abundance mole fractions include: Fischer-Tropsch to non-catalytic adenine, Fischer-Tropsch to non-catalytic guanine and Fischer-Tropsch adenine to reaction 62 of thymine (NC). Corresponding initial limiting reagent concentration ratios are: NH$_3$ to HCN, NH$_3$ to HCN and NH$_3$ to H$_2$CO, respectively. The values are calculated from initial concentrations in Table~\ref{InitialConcentrations} and simulation abundances of reactions 1, 3, 51, 54 and 62 in Figures~\ref{AdenineSim} and \ref{GuaThySim}.\label{LimitingReagent}}
\begin{tabular}{llcc}
\\
\multicolumn{1}{l}{Numerator} &
\multicolumn{1}{l}{Denominator} &
\multicolumn{1}{c}{Nucleobase Mole Ratio} &
\multicolumn{1}{c}{Limiting Reagent Mole Ratio}\\ \hline
FT Adenine & NC Adenine & 2.8 & 2.8 \\
FT Guanine & NC Guanine & 2.8 & 2.8 \\
FT Adenine & NC Thymine* & 4.2 & 3.8 \\
\hline
\multicolumn{4}{l}{\footnotesize NC: Non-catalytic, CA: Catalytic, FT: Fischer-Tropsch.}\\
\multicolumn{4}{l}{\footnotesize *Reaction 62}
\end{tabular}
\end{table*}

The crucial finding is that all production ratios match their corresponding limiting reagent ratios very well. This means that the production of these nucleobase reactions is mainly driven by the initial concentration of each reaction's limiting reagent.

\subsection{Why Isn't Thymine Found in Observed Meteorites?}

As we have shown in the previous section, nucleobase synthesis within planetesimals is mainly driven by the initial limiting reagent abundances for each reaction. Formaldehyde, the limiting reagent of glycine synthesis within meteorite parent bodies \citep{Reference44}, is also the limiting reagent of reaction 62, the only favourable reaction to produce T in our simulations. Therefore since glycine is the most abundant proteinogenic amino acid measured in carbonaceous chondrites, it is unlikely that T synthesis within meteorite parent bodies is affected by a scarcity of formaldehyde. Furthermore, competition for reactants between the Strecker synthesis of glycine and reaction 62 of T appears to only reduce T synthesis between $\sim$0--25$^{\circ}$C (see Figure~\ref{GuaThySim}). Instead, we propose that a unique decomposition pathway is disallowing T to persist through the aqueous stage of meteorite parent body interiors.

Laboratory experiments by \citet{Reference55} have shown that T decomposes by 18$\%$ in just 40 minutes when heated to 120$^{\circ}$C in an aqueous solution of hydrogen peroxide (H$_2$O$_2$). H$_2$O$_2$ is found in the spectrum of comet Hale-Bopp in abundances of $\sim$0.03 mol/100 mol H$_2$O \citep{2004A&A...418.1141C}. Therefore it is conceivable that H$_2$O$_2$ was incorporated into meteorite parent bodies at the time of the latter's formation. Since 120$^{\circ}$C is within the likely range of temperatures within carbonaceous chondrite parent bodies, any T produced from reaction 62 within these bodies could have been quickly decomposed by H$_2$O$_2$. The interesting question is: do most planetesimals incorporate hydrogen peroxide during their formation?

Unfortunately, due to lack of experimentation, it is unknown whether H$_2$O$_2$ decomposition is a selective process for T, or if the former can also decompose the other four nucleobases. Therefore our hypothesis for the apparent lack of T within carbonaceous chondrites still requires further experimental validation.

\subsection{Simulated Nucleobase Abundances vs. Meteoritic Abundances}

An important discrepancy arises between the nucleobase abundances from our individual FT and NC simulations and the nucleobase abundances in carbonaceous chondrites. Our simulations produce 3--4 orders of magnitude more nucleobases than are present in the meteoritic record. This could be expected, as all nucleobases decay due to hydrolysis at a rate that increases with temperature \citep{Reference45}. For example, in an aqueous environment, G deaminates into xanthine and A into hypoxanthine, however experiments have demonstrated that these purine nucleobases are less susceptible to deamination than cytosine deamination \citep{Reference67}. One of the caveats of using an equilibrium chemistry model is that we cannot simulate the decomposition of nucleobases if the decay rate exceeds the time the model planetesimal has to reach equilibrium (in our case, millions of years). At some temperatures G, A, U and T have half-lives in aqueous solution of $\geq$ 10$^6$ years \citep{Reference45}, therefore only the decomposition of C (max half-life: $\sim$17,000 years) could be included in our simulations (reaction no. 32). This limitation results in simulated abundances of G, A, U and T that are higher than expected.

The high (7.1 wt$\%$) water content within our model planetesimal can also help explain the superfluous individual nucleobase simulation abundances with respect to the meteoritic record. Models with one two-thousandth the fiducial planetesimal water content (by volume) have shown that A, G and U simulation abundances can fall into the range of meteoritic abundances of A, G and U. Though it is unlikely that carbonaceous chondrite parent bodies had only 0.003 wt$\%$ water (measurements of petrographic type 1--3 carbonaceous chondrites have revealed water contents in the range 0.3--22 wt$\%$ \citep{2006mess.book...19W}), reducing the water content within our model planetesimal would contribute to reducing the nucleobase abundances in our individual reaction simulations.

Besides nucleobase decay and water content, it is also important to note that molecular competition isn't considered in our individual reaction simulations. As shown in Figures~\ref{AdenineSim} and \ref{GuaThySim}, the NC reactions of A, G and T all decrease in production when competing with the Strecker synthesis of glycine for reactants. This effect even causes reaction 58 of T to be unproductive at temperatures $<$ 200$^{\circ}$C. A similar effect is found when each of the five NC nucleobase reactions from a theoretical study \citep{2008OLEB...38..383L} and an additional NC thymine reaction \citep{Reference53} are simulated together (Figure~\ref{LaRoweSim}). All six nucleobase reactions decrease in production when competing with each other for reactants. Though the results of this competition simulation are likely less accurate than the individual simulation results (see Section~\ref{accuracy}), this simulation at least shows the potential for a reduction in nucleobase synthesis due to the mutual competition for reactants.

A final consideration for the high production of nucleobases in our simulations with respect to the meteoritic record is the potential nucleobase decay during atmospheric entry. Since nucleobases decay rapidly from hydrolysis at higher temperatures, if the interior of a meteorite were to reach temperatures as high as 500$^{\circ}$C, most (if not all) of the nucleobases would decompose during the 5--15 seconds of atmospheric decent. We could assume nucleobase-dissociating temperatures aren't reached within meteorites upon entry simply due to the fact that nucleobases are found in measurable quantities in carbonaceous chondrites, and are strongly thought to be extraterrestrial in origin \citep{2011PNAS..10813995C}. However, additional evidence comes from a model of heat-diffusion in meteorites during atmospheric entry, which reveals that temperatures near 700 K penetrate only as deep as 0.5--1 cm for carbonaceous chondrites \citep{2014arXiv1412.5134S}. This small layer for which temperatures near 700 K can reach---for larger-radius meteorites---composes only a tiny proportion of the organic content.

\subsection{Simulated Relative Frequencies vs. Meteoritic Frequencies}\label{discrepancy}

Both the relative NC nucleobase simulation abundances and the relative FT nucleobase simulation abundances give the same results: the deamination of C into U should result in a dominant abundance of U within meteorite parent bodies. This result is echoed in the competition simulation, where, after considering deamination, U is also the most productive nucleobase at all temperatures. These simulation results significantly differ from observations, as U is actually the third most abundant nucleobase in meteorites after G and A \citep{Reference46}. Therefore we speculate that additional nucleobase decomposition pathways (e.g. the oxidation of C into 5-hydroxyhydantoin \citep{Reference46}) are playing a role in limiting meteoritic U abundances.

\subsection{Regulating FT Synthesis in Carbonaceous Chondrites}\label{ammonia}

Our simulations verify NH$_3$ to be the limiting reagent of FT synthesis (see Section~\ref{adenine}) based on initial concentrations that are dominant in CO and H$_2$ (see Table~\ref{InitialConcentrations} for values). However, CO can also limit FT synthesis as it is the only carbon source in FT reactions. To verify this, we adjust the initial CO concentration to match the initial NH$_3$ concentration (0.7 mol/mol H$_2$O) while leaving the H$_2$ abundance at 17.5 mol/mol H$_2$O, and we rerun each FT synthesis simulation. The results show both NH$_3$ and CO to be equally depleted at equilibrium. (We also rerun FT synthesis simulations with initial CO concentrations which are less than the initial NH$_3$ concentration, and verify that CO would become completely depleted while some NH$_3$ remained.) This means NH$_3$ is only the limiting reagent in FT synthesis when the initial molar NH$_3$:CO concentration ratio is less than 1:1. This is the case for our model planetesimal, where NH$_3$:CO is 0.04:1, and the laboratory experiments by \citet{1968GeCoA..32..175H}, where NH$_3$:CO ranged from 0.15--0.6:1. 

If there is a reaction that is competing with FT synthesis for the CO reactant within meteorite parent bodies, then the effective NH$_3$:CO ratio for FT synthesis may reach above 1:1. In this case, we would expect nucleobase production to decrease due to the depletion of usable CO for FT synthesis within these parent bodies. This would also result in some leftover NH$_3$, as in CO-deplected parent bodies, NH$_3$ would no longer be the limiting reagent of FT synthesis.

This could be what happened in CR2 meteorite parent bodies, whose meteorites have some of lowest abundances of total nucleobases, at 6--25 ppb \citep{2011PNAS..10813995C}, and the highest abundances of NH$_3$, at $\sim$14--19$\mu$mol/g \citep{2009GeCoA..73.2150P}. Conversely, CM2 meteorites, which have lower abundances of NH$_3$, at $\sim$0.3--1.1$\mu$mol/g \citep{2011GeCoA..75.7585M}, and the highest abundances of total nucleobases, at 22--788 ppb \citep{2011PNAS..10813995C,Reference37,1981GeCoA..45..563S,1979Natur.282..709S,1977GeCoA..41..961V}, may have had a less efficient depletion of CO within their parent bodies, allowing FT synthesis to produce more nucleobases and leave behind less NH$_3$. The lack of CO found within carbonaceous chondrites supports a CO limiting reagent for FT synthesis within their parent bodies. In the case of CR2 meteorites, where FT synthesis may be the most curbed by a competing reaction, NC synthesis has the potential to produce a more significant fraction of the nucleobases within CR2 meteorite parent bodies.

One possibility for a reaction that competes against, and thus regulates FT synthesis might be CO$_{(g)}$ + H$_2$O$_{(l)} \rightarrow$ Formic acid. This reaction is simple, and other than CO, requires only liquid water which is abundant in carbonaceous chondrite parent bodies. The product of this reaction, formic acid, has also been measured in both CM and CV meteorites \citep{1993Metic..28..330B}, which increases the possibility of this reaction being a valid competitor.

\subsection{Most Important Reactions}

In Table~\ref{MostImportant} we summarize our findings by listing the most important nucleobase synthesis reactions within planetesimals and their corresponding simulation abundances at 100$^{\circ}$C and 100 bar. Two reactions are displayed for each nucleobase, which correspond to the most likely candidates to have produced nucleobases within meteorite parent bodies. We see that these are either FT or NC reactions. Only one NC reaction was chosen for A synthesis (no. 3) to represent all similar and equally-productive A reactions of this type (nos. 4, 7, and 8). Our simulations have shown that FT synthesis tends to produce a factor of 2 to 4 more nucleobases within planetesimals than NC synthesis, however NC synthesis should not be neglected. The nucleobase analogs and catabolic intermediates found within carbonaceous chondrites (purine, 2,6-diaminopurine, 6,8-diaminopurine, xanthine, and hypoxanthine) cannot be produced by any known FT reaction pathway, but are co-products with A and G in laboratory experiments demonstrating NC synthesis \citep{2011PNAS..10813995C}. This provides evidence that NC and FT synthesis are likely occurring in parallel within meteorite parent bodies.
The most important reactions involve simple molecules such as HCN, CO, NH$_3$ and water. These are ultimately supplied by the protoplanetary disk out of which planetesimals were formed. In this sense nucleobase synthesis, as well as that of amino acids, is tightly coupled to the astrochemistry of protoplanetary disks.

\begin{table*}[t]
\centering
\caption{Abundances produced by the most important individual nucleobase reaction simulations at 100$^{\circ}$C and 100 bar. These reactions represent the candidates that most likely produced nucleobases within meteorite parent bodies. Only one NC reaction of A synthesis (no. 3) is displayed to represent all similar and equally-productive A reactions of this type (nos. 4, 7, and 8). The reaction equations are the same as those from Table~\ref{Candidates}. \label{MostImportant}}

\begin{tabular}{lclc}
\\
\multicolumn{1}{l}{No.} &
\multicolumn{1}{c}{Type} &
\multicolumn{1}{l}{Reaction} &
\multicolumn{1}{c}{Abundance (x10$^5$ ppb)}\\ \hline
$\underline{Adenine}^a$ & \\
1 & FT & CO + H$_{2}$ + NH$_{3}$ $\xrightarrow{NiFe+||Al_{2}O_{3}+||SiO_{2}}$ A + H$_{2}$O & 7.37\\
3 & NC & 5HCN$_{(aq)}$ $\rightarrow$ A$_{(aq)}$ & 2.66\\
$\underline{Uracil}^b$ & \\[+1mm]
29 & NC & 2HCN$_{(aq)}$ + 2CH$_{2}$O$_{(aq)}$ $\rightarrow$ U$_{(aq)}$ + H$_{2(aq)}$ & 1.46\\
32 & NC & C + H$_{2}$O $\rightarrow$ U + NH$_{3}$ & 10.25$^d$\\
$\underline{Cytosine}^b$ & \\
43 & FT & CO + H$_{2}$ + NH$_{3}$ $\xrightarrow{NiFe+||Al_{2}O_{3}+||SiO_{2}}$ C + H$_{2}$O & 10.14\\
44 & NC & 3HCN$_{(aq)}$ + CH$_{2}$O$_{(aq)}$ $\rightarrow$ C$_{(aq)}$ & 2.89\\
$\underline{Guanine}^c$ & \\
51 & FT & CO + H$_{2}$ + NH$_{3}$ $\xrightarrow{NiFe+||Al_{2}O_{3}+||SiO_{2}}$ G + H$_{2}$O & 8.26\\
54 & NC & 5HCN$_{(aq)}$ + H$_{2}$O $\rightarrow$ G$_{(aq)}$ + H$_{2(aq)}$ & 2.98\\
$\underline{Thymine}^c$ & \\[+1mm]
58 & NC & 2HCN$_{(aq)}$ + 3CH$_{2}$O$_{(aq)}$ $\rightarrow$ T$_{(aq)}$ + H$_{2}$O & 1.09\\
62 & NC & U + CH$_{2}$O + Formic Acid + H$_{2}$O $\rightarrow$ T & 1.64\\
\hline
\multicolumn{4}{l}{\footnotesize $^a$Adenine simulation results are from Figure~\ref{AdenineSim}} \\
\multicolumn{4}{l}{\footnotesize $^b$Uracil and Cytosine simulation results are from Figure~\ref{CytUraSim}} \\
\multicolumn{4}{l}{\footnotesize $^c$Guanine and Thymine simulation results are from Figure~\ref{GuaThySim}} \\
\multicolumn{4}{l}{\footnotesize $^d$Cytosine abundance from reaction 43 used as input cytosine concentration for this reaction.} \\
\multicolumn{4}{l}{\footnotesize NC: Non-catalytic, CA: Catalytic, FT: Fischer-Tropsch.} \\
\end{tabular}
\end{table*}

\section{Conclusions}\label{conclusionssec}

We summarize our most important findings below.

\begin{itemize}
\item Our thermochemical simulations reveal that cytosine (C) is unlikely to persist within meteorite parent bodies as it efficiently decomposes in water to produce uracil (U) and NH$_3$. This reaction has a half-life of less than 17,000 years \citep{Reference45}, which is at least 100 times less than the period for which planetesimals are thought to have had aqueous interiors \citep{2005E&PSL.240..234T}. 

\item Our simulations show that thymine (T) has a thermodynamically favourable reaction pathway from U, formaldehyde and formic acid. Though T, like C, is also unlikely to persist within meteorite parent bodies (this time due to an efficient oxidation reaction with H$_2$O$_2$). H$_2$O$_2$ has been shown to decompose 18$\%$ of aqueous T in 40 minutes at 120$^{\circ}$C \citep{Reference55}. And since H$_2$O$_2$ has been found in the spectra of comets  \citep{2004A&A...418.1141C}, it is conceivable that H$_2$O$_2$ was available to oxidize T in at least the parent bodies of the carbonaceous chondrite meteorites that we have today.

\item Individual FT reactions produce nucleobase abundances in the range of 7--10x10$^5$ ppb, and most individual NC reactions produce abundances in the range of 1--3x10$^5$ ppb. For each individual nucleobase simulation, FT synthesis tends to produce a factor of 2 to 4 more nucleobases within planetesimals than NC synthesis. This suggests that FT synthesis is the most prominent reaction-type for nucleobase formation within most planetesimals.

\item NC synthesis likely produces a more significant fraction of the nuclebases within CR2 meteorite parent bodies. Evidence for this is in the high abundances of the usual limiting reagent of FT synthesis, NH$_3$, measured within CR2 meteorites ($\sim$14--19$\mu$mol/g) \citep{2009GeCoA..73.2150P}. NH$_3$ would likely only remain within CR2 meteorites if CO, the only carbon source in FT synthesis, was efficiently depleted within CR2 parent bodies. The efficient depletion of CO within CR2 parent bodies would curb FT synthesis, allowing NC synthesis to contribute a greater fraction of the total nucleobase inventory within these bodies.

\item The deamination of C into U is the most abundant NC reaction, as it produces an equivalent amount of U as the input C concentration.

\item Simulating nucleobase reactions while also allowing the Strecker synthesis of glycine to occur has shown a decrease in A, G and T production within our model planetesimal. Allowing five NC nucleobase reactions from a theoretical study \citep{2008OLEB...38..383L} and an additional NC thymine reaction \citep{Reference53} to react in a single simulation has shown a decrease in the production of all nucleobases with respect to their individual reaction simulations. Molecular competition for reactants, the high water content in our model planetesimal, and the decomposition of nucleobases due to hydrolysis \citep{Reference45} can help explain the lower levels of nucleobases in carbonaceous chondrites with respect to the individual nucleobase simulations.

\item The relative simulation abundance of A to G for FT synthesis is 1. If we consider that the deamination of C into U simulation produces an amount of U equivalent to the input concentration of C, the relative U to G, originating from the FT synthesis of C, is 1.67. The relative simulation abundance of A to G for NC synthesis is also 1. If we calculate the relative U to G abundance for NC synthesis using the sum of the U and C produced from NC synthesis---again because C will completely deaminate into U at equilibrium---then the relative U to G is 1.97. The relative A to G simulation abundances are equivalent for FT and NC synthesis, and are slightly outside of the error of the relative A to G within carbonaceous chondrites of 0.36 $\pm$ 0.48. The relative U to G simulation abundances for FT and NC synthesis are also similar, but are much higher than the relative U to G within carbonaceous chondrites of 0.23 $\pm$ 0.19 \citep{Reference46}.

\item The large discrepancy between the relative simulation abundances of U to G compared to the meteoritic record may hint at the importance of another decay mode occuring within planetesimals, such as the oxidation of C into 5-hydroxyhydantoin \citep{Reference46} (which would offer less C to deaminate into U).

\item Finally, the approximate temperature limit for effective nucleobase synthesis within planetesimals is 165$^{\circ}$C.
\end{itemize}

There are several broad and interesting questions that follow from these results. If, as our calculations suggest, T and C are truly absent from all meteorites, then the origin of the materials for building the genetic code takes on a very interesting new twist. Our reactions are hydrothermal in character therefore some of the constraints we have found would also pertain to hydrothermal systems on planets. One possibility is a cometary source for C and T. Recent experiments have produced C and trace amounts of T on icy grains \citep[and references therein]{2014ApJ...793..125N}. A different possibility is that nature produced nucleotides directly, as \citet{Reference56} have proposed, bypassing the need to produce nucleobase "lego blocks". Unlike the amino acids and protein synthesis, the ultimate question of RNA and DNA synthesis may involve greater complexity unless it can be shown that a robust chemistry for nucleotide synthesis on early planets exists. 

Perhaps the most conservative and interesting possibility is that the earliest form of the RNA world could function with only 3 nucleobases supplied by meteorites. Functional riboyzomes with only A, G, and U have been made in the laboratory \citep{Reference65}. In this event, we speculate that it may be possible that meteorites supplied a minimal but still realizable set of molecules for the establishment of a precursor RNA world. We plan to investigate these questions in future papers.

\section*{Acknowledgements}

We thank the anonymous referees for reports that helped to improve this paper. We are indebted to Alyssa Cobb, who laid the groundwork for this research
by investigating amino acid synthesis within meteorite parent bodies. We are also extremely thankful to Anand Patel, Paul Ayers, and Farnaz Zadeh for helping verify our substitution reactions and for sharing their chemistry knowledge and expertise with us. The research of B.K.D.P. was supported by a NSERC CREATE Canadian Astrobiology Training Program Undergraduate Fellowship as well as an NSERC postgraduate scholarship at McMaster University (Canada Graduate Scholarship-Master's). R.E.P. is indebted to the Institut f{\"u}r Theoretische Astrophysik and the Max-Planck Institut f{\"u}r Astronomie in Heidelberg for support of his sabbatical leave where work on this manuscript was completed. R.E.P. is supported by an NSERC Discovery Grant.

\section*{Author Disclosure Statement}

No competing financial interests exist. This is a copy of an article published in Astrobiology 2016. Astrobiology is available online at \href{http://online.liebertpub.com}{http://online.liebertpub.com}.

\section*{Appendix A - CA Synthesis}

Our simulations of CA synthesis (reaction nos. 24 of A, 49 of C, 62 of U, and 63 of T) are unproductive at all temperatures. Due to an absence of Gibbs free energy data for the liquid formamide reactant, these simulations utilize the Gibbs free energies of a structurally similar, substitute reactant: the carbamoyl functional group (-CONH$_2$). Though these two molecules are similar in steric configuration, they are bound to have some small variation in their Gibbs free energies of formation. To quantify the difference in Gibbs free energies between the substitute reactant and formamide, we consulted the Ayers group in the Department of Chemistry and Chemical Biology at McMaster University. Patel et al. (2015, personal communication) estimated the difference in Gibbs free energies between these molecules using an electronic structure modeling program called Gaussian. Their analysis showed that the difference in Gibbs free energies of formation between the aqueous carabamoyl functional group and liquid formamide is likely only 25 kJ/mol, with formamide having the lower value. Overall, the 25 kJ/mol Gibbs free energy difference should be considered a close approximation.

%However there is likely a slight error in this value, associated with additional, unaccounted for solvent-solute interactions.

To see if 25 kJ/mol is a large enough difference in Gibbs free energy to affect our unproductive simulations of CA synthesis, we decrease the Gibbs free energy of formation of the substitute reactant by this amount for all temperatures, and rerun each CA synthesis simulation with the adjusted molecular data. These simulation reruns of CA synthesis are also unproductive at all temperatures. (Note that these results could have also simply been assumed, as lowering the Gibbs free energy of a stable reactant makes the reactant even more likely to remain itself than react and create a product.)

%It is also worth noting that formamide has a relatively low abundance in comets (0.015 mol formamide/100 mol H$_2$O) compared to the cometary abundances of the reactants in NC and FT synthesis (e.g. 0.25 mol HCN/100 mol H$_2$O, 17.5 mol CO/100 mol H$_2$O). Formamide's low abundance in comets was conceivably similar to formamide's abundance within carbonaceous chondrite parent bodies. Thus CA synthesis is unlikely to produce abundances of A, C, U or T as high as those produced from NC and FT synthesis.

\section*{Appendix B - Caveats in Modeling Competition Between Reactions}\label{compete}

When modeling several nucleobase reactions at once, our resultant abundances tend to disagree with the experimental results in the lab. This is demonstrated most obviously by the lack of A produced at low temperatures in the competition simulation between 6 NC nucleobase synthesis reactions (Figure~\ref{LaRoweSim}), and the lack of G and A produced in the competition simulation between the 3 FT nucleobase synthesis reactions.
The likely cause of these disagreements is that our competition models are too constrained. In our models, we only allow certain reactions to occur by restricting the reactants and allowed products. In the lab, there are no such restrictions, and nucleobase synthesis will compete with several reactions simultaneously. For example, it has been demonstrated that the formation of 8-hydroxymethyladenine is favoured over A for increased formaldehyde concentrations \citep{Reference62}. On the other hand, formaldehyde reacts rapidly in solutions with HCN present, forming cyanohydrins, the latter of which accelerates the rate of HCN oligomerization and thus A synthesis \citep{Reference63,Reference12}. This tension between the promotion and depression of A synthesis is just one example of the complex competition between reactions that could be occurring within meteorite parent bodies. Thus including all of the necessary constraints is essential in order to truly simulate the complexity of competition between reactions using equilibrium chemisty models.

\bibliography{Bibliography}

\end{document}